\documentclass[preprint,authoryear,12pt]{elsarticle}
\usepackage{graphics}
\usepackage{amssymb}
\usepackage[nodots]{numcompress}
\journal{MNRAS}

\def\lsim{\ ^{<}\!\!\!\!_{\sim}\>}
\def\gsim{\ ^{>}\!\!\!\!_{\sim}\>}
       
\begin{document}

\begin{frontmatter}

\title{The end states of long-period comets and the origin of Halley-type comets}


\author{Julio A. Fern\'andez, Tabar\'e Gallardo, Juan D. Young}
%\ead{gallardo@fisica.edu.uy}

\address{Departamento de Astronom\'{i}a, Facultad de Ciencias, Universidad de la Rep\'ublica, Igu\'{a} 4225, 11400 Montevideo, Uruguay}

\begin{abstract}
We analyze a sample of 73 old long-period comets (LPCs) (orbital periods $200 < P < 1000$ yr) with perihelion distances $q < 2.5$ au, discovered in the period 1850-2014. We cloned the observed comets and also added fictitious LPCs with perihelia in the Jupiter's zone. We consider both a purely dynamical evolution and a physico-dynamical one with different physical lifetimes. We can fit the computed energy distribution of comets with $q < 1.3$ au to the observed one only within the energy range $0.01 < x < 0.04$ au$^{-1}$ (or periods $125 < P < 1000$ yr), where the ``energy'' is taken as the inverse of the semimajor axis $a$, namely $x \equiv 1/a$. The best results are obtained for physical lifetimes of about 200-300 revolutions (for a comet with a standard $q = 1$ au). We find that neither a purely dynamical evolution, nor a physico-dynamical one can reproduce the long tail of larger binding energies ($x \gsim 0.04$ au$^{-1}$) that correspond to most Halley-type comets (HTCs) and Jupiter-family comets. We conclude that most HTCs are not the end states of the evolution of LPCs, but come from a different source, a flattened one that we identify with the Centaurs that are scattered to the inner planetary region from the trans-Neptunian belt. These results also show that the boundary between LPCs and HTCs should be located at an energy $x \sim 0.04$ au$^{-1}$ ($P \sim 125$ yr), rather than the conventional classical boundary at $P = 200$ yr.

\end{abstract}


\end{frontmatter}

\bigskip

\section{Introduction}
\label{intr}

\begin{figure}[h]
\resizebox{10cm}{!}{\includegraphics{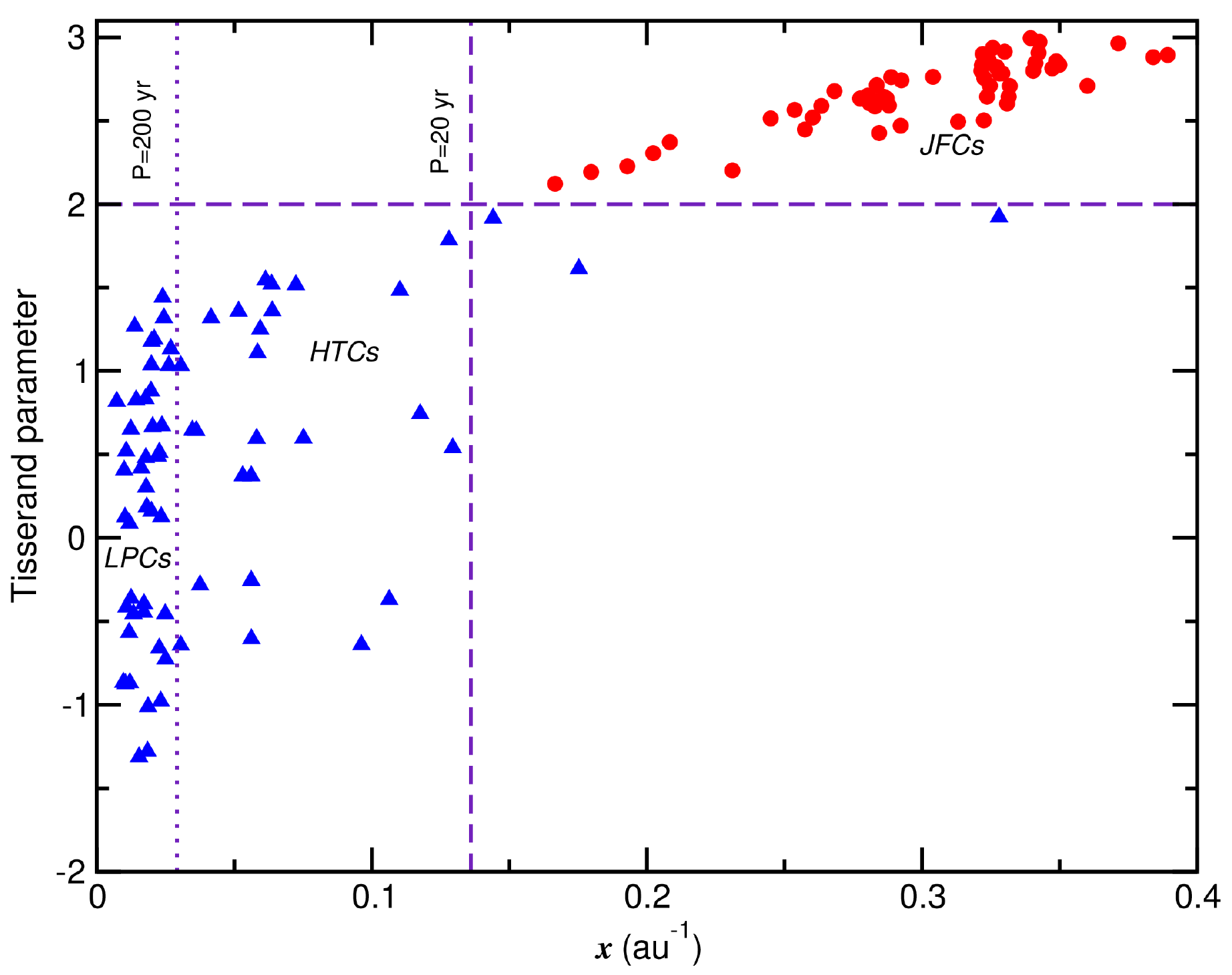}}
\caption{Tisserand parameter versus the orbital energy of old LPCs ($200 < P < 1000$ yr) and HTCs (blue triangles), and JFCs (red circles) with perihelion distances $q < 1.3$ au (Source: {\it IAU Minor Planet Center www.minorplanetcenter.net}).}
\label{tisserand}
\end{figure}

The origin of the Halley-type comets (HTCs) has been a matter of debate since their orbital properties place them as an ``intermediate'' population between long-period comets (LPCs) and Jupiter family comets (JFCs). In Fig. \ref{tisserand} we show the Tisserand parameters with respect to Jupiter of the observed populations of 'old' LPCs (orbital periods $200<P<10^3$ yr), HTCs and JFCs with perihelion distances $q < 1.3$ au as a function of their orbital energies. We note that the ``energy'' is represented by the variable $x \equiv 1/a$, since the orbital energy is proportional to the inverse of the semimajor axis $a$. The Tisserand parameters with respect to Jupiter of HTCs:$T_J < 2$ share the same range as the LPCs, whereas JFCs seem to occupy a different niche with $T_J > 2$.\\

The inclination distribution of HTCs shows a strong predominance of direct orbits, while LPCs show a more or less isotropic distribution of inclinations. Because HTCs have shorter orbital periods than LPCs, the former are prone to fall in mean-motion resonances with Jupiter, as for instance 1:5, 1:6, 1:7 which can help to increase their dynamical stability \citep{Caru87, Bail96}. The dynamical lifetime of HTCs is found to be of the order of $10^5-10^6$ yr, and collisions with asteroids do not play a significant role in constraining their lifetime \citep{Vand12}. Yet, physical processes like sublimation, outbursts and splittings may play a much more fundamental role in setting an upper limit to the comet's lifetime in the near-Earth region.\\

It is a well established observational feature that old LPCs are very scarce in the near-Earth region, and that they are almost absent for $q \gsim 2$ au, in contradiction with what should be expected from dynamical grounds, namely a uniform distribution of orbital energies (which would imply a large population of old LPCs). This discrepancy has been attributed to fading (e.g. Neslu$\breve{s}$an 2007), a long discussion that goes back at least to \citet{Oort51}. It is very likely that most comets during their evolution from 'new' to an old LPC lose material via sublimation, outbursts, fragmentation and their ultimate disruption \citep{Levi02}. From numerical simulations, \citet{Wieg99} found that the computed distribution of LPCs could be matched to the observed one by applying a fading law $\propto m^{-0.6 \pm 0.1}$, where $m$ is the number of apparitions.\\

LPCs are thought to come from the Oort cloud, in a dynamical process in which they change their orbital energy in every perihelion passage in a random way until either they are ejected to interstellar passage, or they become more tightly bound to the Sun in orbits of smaller orbital periods \citep{Fern81}. In this scenario we may argue that 'old' LPCs, and even the HTCs, represent dynamically evolved comets with many perihelion passages by the inner planetary region since their insertion from the Oort cloud. Describing the orbit evolution of the comet by a random-walk in the energy space, we find that a new comet injected in the near-Earth region for the first time will require a few hundred revolutions to reach the 'old' state (e.g. Fern\'andez 2005). A finite physical lifetime due to sublimation, outbursts and splittings may account for the predominance of direct orbits, since they require less revolutions to reach a HTC orbit \citep{Fern94}. According to this view, LPCs and HTCs would represent the same dynamical class of comets, being the boundary at $P=200$ yr conventional, since 'periodic' comets with periods shorter than 200 yr have had the chance to be observed more than once.\\

The idea that HTCs are captured from the Oort cloud and thus represent the end state of the dynamical evolution of LPCs was further developed by \citet{Emel98}, \citet{Fern99}, \citet{Wieg99}, \citet{Nurm02} and \citet{Fouc14} among others. From numerical simulations of $10^5$ fictitious Oort cloud comets with initial perihelion distances spread between 0 and 31 au, \citet{Emel98} found that most HTCs originate from LPC orbits with $q < 2$ au, though they also concluded that Oort cloud comets captured in the outer solar system may constitute a significant additional source of HTCs. \citet{Nurm02} also found that HTCs can be captured from the classical Oort cloud, getting good fits for the inclination and Tisserand parameter distributions, and a good agreement with the observed number of comets when physical evolution was included. The problem is that they used in their simulations an \citet{Arno65} type algorithm, in which only close encounters with the planets are considered, thus neglecting the weak perturbations from more distant encounters. The Arnold scheme enhances the capture rate into strongly bound orbits (HTCs and short-period orbits), so the capture rate from Oort cloud comets to HTCs could have been overestimated in their model.\\

Yet, the view that HTCs are the final product of the dynamical evolution of LPCs has been challenged by other authors. \citet{Olss88} argued that 1P/Halley and other intermediate-period comets on retrograde orbits could have been captured from orbits with perihelia near Neptune, semimajor axes $\lsim 100$ au, and inclinations near $60^{\circ}$. After a close encounter with Neptune, there was a perihelion - aphelion exchange, leading to a new retrograde orbit with a small perihelion distance. Even though we consider extremely improbable such dynamical mechanism, Olsson-Steel's idea was interesting in the sense that his source region can be identified in current terms with the scattered disk. From massive integrations of a large number of test particles coming form the Oort cloud in isotropic orbits, \citet{Levi01} found that they could not reproduce the observed inclination distribution of HTCs. They could obtain a good match if they added an extra disk-like inner cloud to the outer isotropic Oort cloud. Later, \citet{Levi06} reviewed their previous model and concluded that a flattened inner Oort cloud cannot be a suitable source, since the orbital inclinations will change by galactic tides much faster than the time scale for decreasing their perihelion distances below $q \simeq 25$ au where their evolution will continue into HTCs. They suggested instead that the Scattered Disk will be the dominant source. Again they argued that SDOs will be scattered by planetary perturbations to Oort cloud distances where galactic tides can drive their perihelia inside the planetary region much faster than the time required to randomize their inclinations and, hence, they can supply a population of mostly low-inclination objets to the HTC population.\\

The previous overview tells us that the source region of HTCs is still controversial. In order to try to elucidate this issue, we carry out numerical simulations partly based on the observed sample of 'old' LPCs with $q < 2.5$ au, complemented with simulations with samples of fictitious comets.

\section{The observed population of old LPCs and HTCs in the near-Earth region}
\label{obse}

Our study will be mainly based on comets reaching the near-Earth region (perihelion distances $q < 1.3$ au) because it is for this region that we have better statistics, and all our comparisons between observed and computed results will be for this range of perihelion distances. Nevertheless, we will also consider the sample of old LPCs with $1.3 < q < 2.5$ au for which the discovery rate has been greatly improving, mainly in the last fifteen years thanks to different sky surveys.\\

\begin{figure}[h]
\resizebox{10cm}{!}{\includegraphics{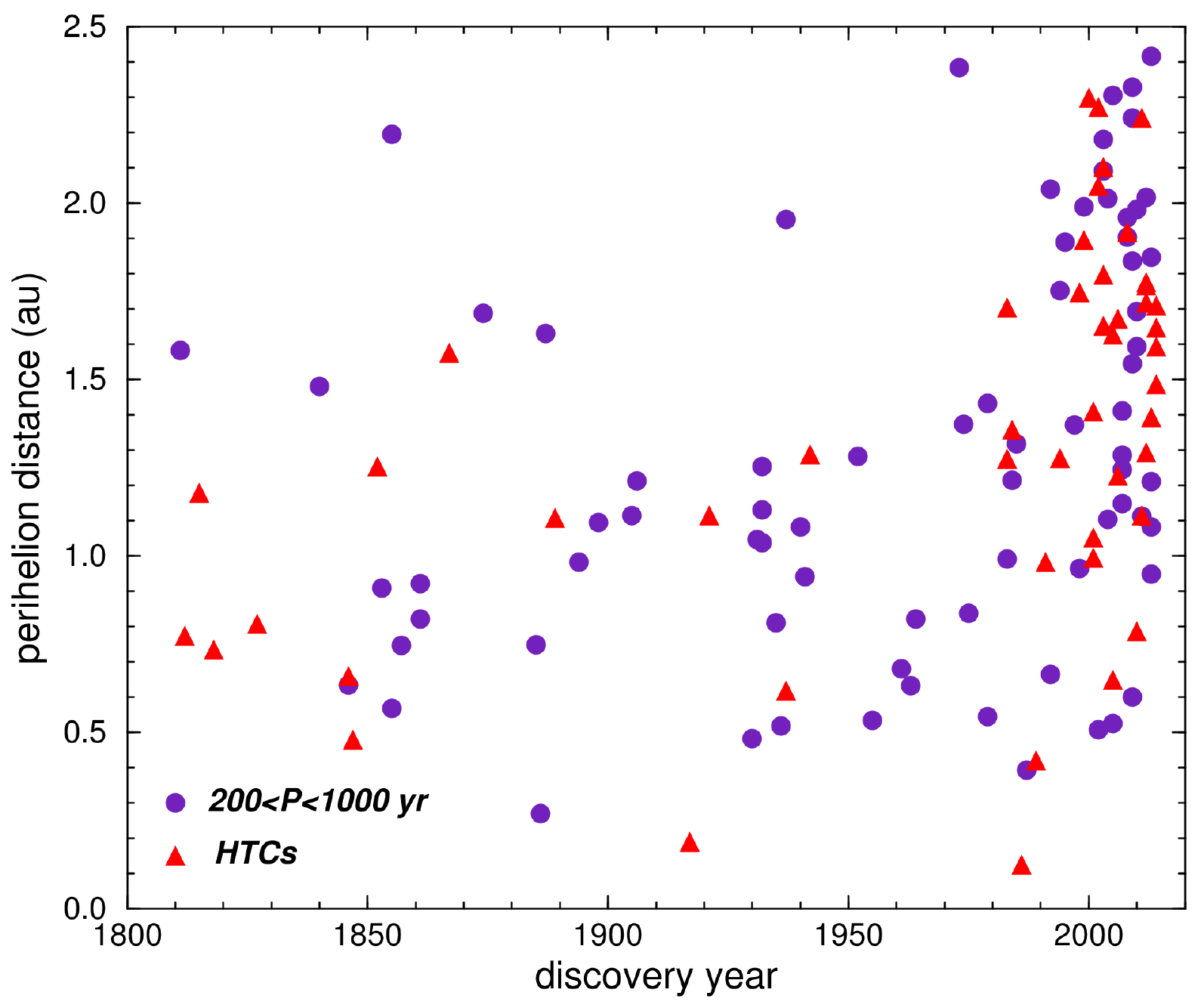}}
\caption{The discovery rate of old LPCs ($200 < P < 1000$ yr) and HTCs with perihelion distances $q < 2.5$ au since 1800 (Source: {\it IAU Minor Planet Center www.minorplanetcenter.net}).}
\label{discovery}
\end{figure}

The record of comet apparitions has been very scarce prior to around 1800, since there were very few experienced observers who could provide an accurate account of the comets' sky positions in order to compute good orbits. The discovery rate increased notoriously during the nineteenth century with the aid of telescopes and, toward the end of the century, the photography. Since around mid-nineteenth century, the discovery rate of comets reaching the Earth neighborhood (perihelion distances $q < 1.3$ au) has not changed too much, as can be seen in Fig. \ref{discovery}.\\

\begin{figure}[h]
\resizebox{10cm}{!}{\includegraphics{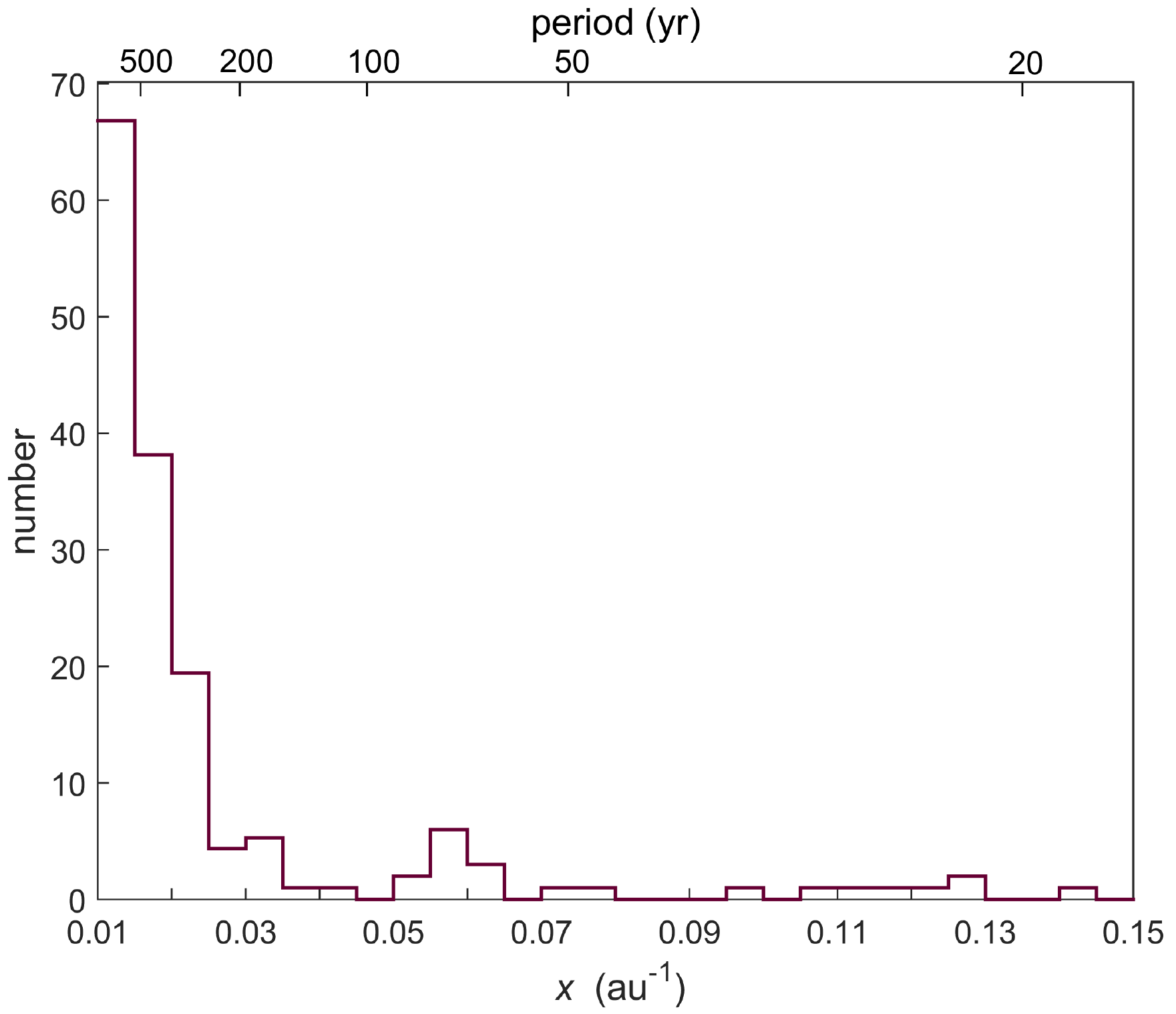}}
\caption{Energy distribution of old LPCs ($200 < P < 1000$ yr) and HTCs observed during 1850-2014 (=164 yr) with perihelion distances $q < 1.3$ au. For comets with orbital periods $P > 164$ yr we have applied a correction factor $P/164$ that allows for those members of longer periods as yet undiscovered because they happen not to pass by the perihelion in the 1850-2014 observing window (Source: {\it IAU Minor Planet Center www.minorplanetcenter.net}).}
\label{energies}
\end{figure}

The distribution of orbital energies of 43 old LPCs discovered in the period 1850-2014 (a time span of 164 yr) and 29 HTCs with $q < 1.3$ au is shown in Fig. \ref{energies}. For comets with periods $P > 164$ yr we have applied a correction factor $P/164$ that takes into account the fraction of comets that still remain unobserved because they have not reached perihelion during the observing window of the last 164 yr.  We can see a steep drop in the number of comets for greater energies, a feature that has been attributed to fading, or other physical effects such as collisions with the Sun, the planets or asteroids.\\

\begin{figure}[h]
\resizebox{10cm}{!}{\includegraphics{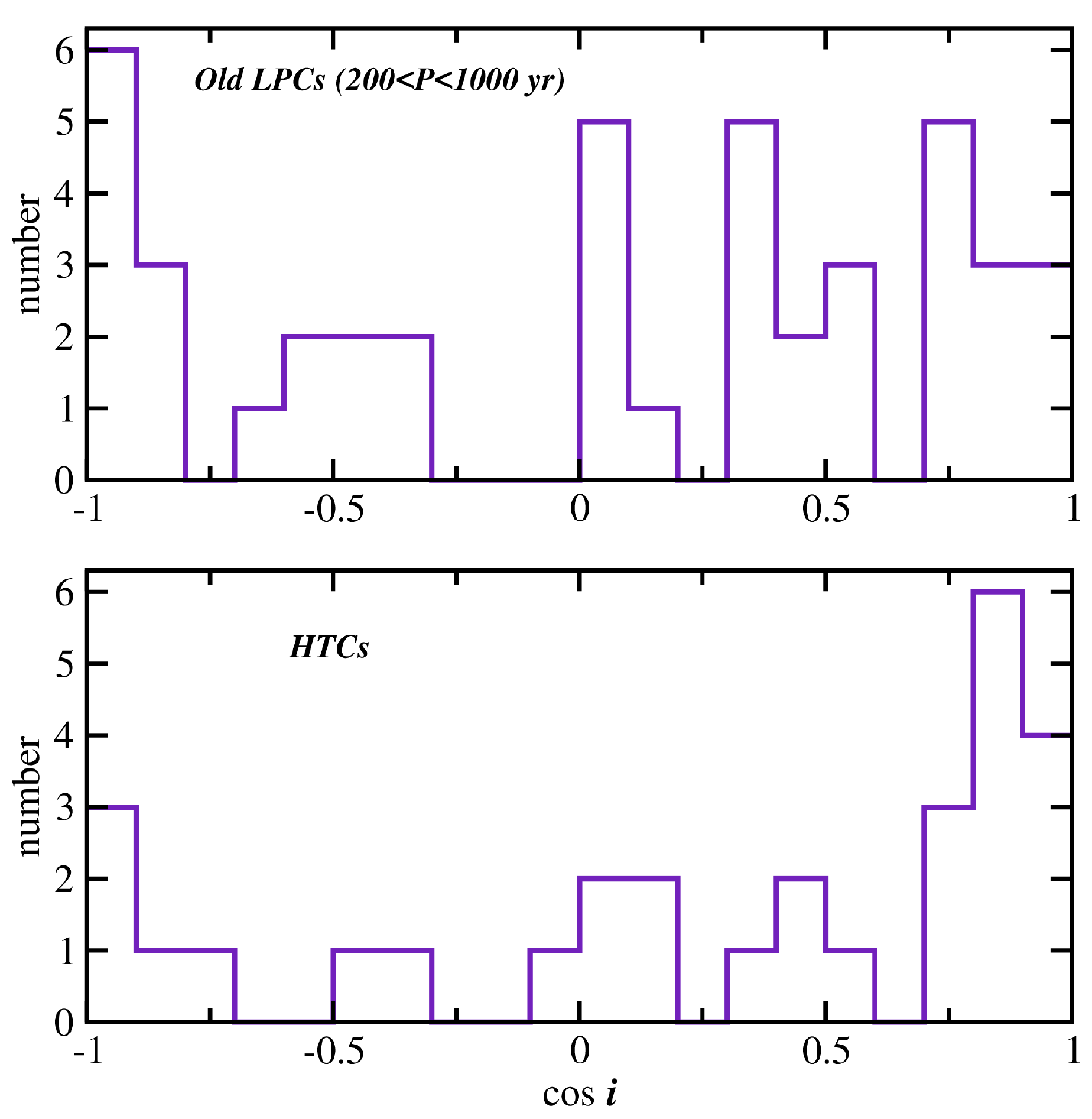}}
\caption{Inclination distribution of old LPCs (upper panel) and HTCs (lower panel) with $q < 1.3$ au observed during the period 1850-2014 (Source: {\it IAU Minor Planet Center www.minorplanetcenter.net}).}
\label{inclinations}
\end{figure}

The inclination distribution (in $\cos{i}$) for old LPCs and HTCs with $q < 1.3$ au show clear departures from randomness (Fig. \ref{inclinations}), that can be seen for instance in the excess of direct orbits (more pronounced for HTCs), and some excess of retrograde comets lying close to the ecliptic plane ($-0.9 > \cos{i} > -1$).\\

As shown before (cf. Fig. \ref{tisserand}), it seems to be a continuation from the LPC population to the HTC population in the parametric plane $(x,T_J)$ that might suggest a common origin. By contrast, the JFC population occupies a different niche with a boundary at $T_J \simeq 2$. There are a few HTCs (with $1.6 \lsim T_J \lsim 2$) that seem to connect both regions, and the question is whether they come from the same source region as the rest of the HTCs and the LPCs. This question will be addressed in this work.

\section{Modelling the dynamical and physical evolution}
\label{mode}

\subsection{The samples}

We consider three samples:

\begin{itemize}

\item 43 'old' LPCs (orbital periods $200 < P < 1000$ yr) discovered between 1850-2014 with perihelion distances $q < 1.3$ au. As said before, we applied a bias correction factor $P/164$. In general for a given comet of period $P$ ($>164$ yr), a weighting factor $w = P/164 \times A$ was applied, where $A$ is a normalization factor to get a number of test objects = 1000. We then generate for a given comet $j$ of period $P_j$ a number of clones $int(w_j)$, i.e. the weighting factor rounded to the closest integer. The clones of a given comet $j$ were taken with random energies and perihelion distances within $\pm 0.5$\% of the energy and perihelion distance of comet $j$, respectively. The orbital angular parameters were taken randomly within $\pm 1$ degree of the corresponding values for comet $j$. In this way we generated a sample S1 of test objects consisting of 43 original LPCs + 957 clones = 1000.

\item 30 'old' LPCs with the same conditions as before but perihelion distances $1.3 \leq q < 2.5$ au. We proceeded as before by applying weighting factors $P/164$, and multiplying all of them by a normalization factor in order to generate 970 clones with orbital elements obtained in the same way as before. We will call S2 to this sample of 30 LPCs + 970 clones.

\item We also generated a sample S3 of 1000 fictitious comets with initial perihelion distances taken randomly within the interval $4 < q < 6$ au, random orientations of their orbital planes, a fixed starting semimajor axis $a = 100$ au, and random values for the other angular orbital elements. The reason why we used here a totally fictitious sample is because of the lack of enough observed old LPCs with perihelia in the region around Jupiter's orbit.

\end{itemize}

In order to check the previous results, we run again the simulations but with three entirely fictitious samples, F1, F2, and F3, of 1000 test bodies each, and chosen within the previously defined $q$ ranges, namely: $q < 1.3$ au (F1), $1.3 < q < 2.5$ au (F2), $4 < q < 6$ au (F3). All the fictitious comets started with a semimajor axis $a = 100$ au, random inclinations $z=\cos{i}$ in the interval $-1 < z < +1$, and argument of perihelion $\omega$ and longitude of the ascending node $\Omega$ also taken at random within the range $(0,2\pi)$. As regards, the perihelion distances, these were taken at random within the three ranges defined above and following the empirical law given by eq.(\ref{qcum}) to be described below.\\

We also added physical effects to the purely dynamical evolution following the procedure that will be described in Section 3.4.\\

Finally, the samples S1, S2 and S3 were run with nongravitational (NG) forces in order to check whether they have any significant influence on the evolution of the comets. For the radial and transverse components of the NG acceleration we used the following expressions

\begin{equation}\label{ngf1}
J_r = A_1 g(r)
\end{equation}

\begin{equation}\label{ngf2}
J_t = A_2 g(r)
\end{equation}
where $g(r)$ is a function of the heliocentric distance $r$ and can be represented by the empirical expression

\begin{equation}\label{ngf3}
g(r) = \alpha \left(\frac{r}{r_o}\right)^{-m} \left[1 + \left(\frac{r}{r_o}\right)^n\right]^{-k}
\end{equation}
where $\alpha=0.1113$ is a normalization factor such that $g(1)=1$, $m=2.15$,
$n=5.093$, $k=4.6142$ and $r_o=2.808$ au \citep{Mars73}. In this analysis we assume that the normal component $A_3=0$. We used the following input parameters: $A_1 =2.0$  au day$^{-2}$, $A_2 = 0.2$ au day$^{-2}$, that are typical for an active LPC. 

\subsection{The variation of the comet flux with the perihelion distance}

We were particularly interested in the population of comets that remained or entered the near-Earth region ($q < 1.3$ au) considering the contribution from the three populations described before. To this purpose we still need to know the relative sizes of the three populations for which it is necessary to estimate how the comet flux varies with the perihelion distance.\\

As mentioned, we considered three comet samples within the ranges of perihelion distances: (S1) $q < 1.3$ au; (S2) $1.3 < q < 2.5$ au; and (S3) $4 < q < 6$ au. If we somehow can define the law $N_c(<q)$, that describes how the cumulative number of old LPCs with perihelion distances $<q$ varies with $q$, we can then estimate the relative sizes of these three populations. The law $N_c(<q)$ can be estimated from the observed flux of old LPCs.\\ 

From the inspection of Fig. \ref{discovery} we find that 11 old LPCs with $q < 1.3$ au, and other 17 with $1.3 < q < 2.5$ au passed by their perihelia during the period 2000-2014. Probably the former sample is near complete, while the latter may be still somewhat incomplete because of their larger distances to the Sun and the Earth. We may also note that the discovery rate of old LPCs in Earth-crossing orbits ($q < 1$ au) shows little change in the considered period 1850-2014 (suggesting a near complete record), and unbiased in $q$. \citet{Fern12} found that the great majority of LPCs with $P>10^3$ yr have absolute total magnitudes $H<12$ (or diameters $D \gsim 0.5$ km). A check of the total absolute magnitudes of the old LPCs of our sample agrees with the previous observation: it is very rare to find comets fainter than $H \sim 12$ among the old LPCs too. Therefore, we will assume in the following that our LPC sample contains comets brighter than $H=12$. Fainter comets either disintegrate very fast, or do not exist by cosmogonic reasons, or are too faint to be detected (see a discussion by Fern\'andez and Sosa 2012).\\

\begin{figure}[h]
\resizebox{12cm}{!}{\includegraphics{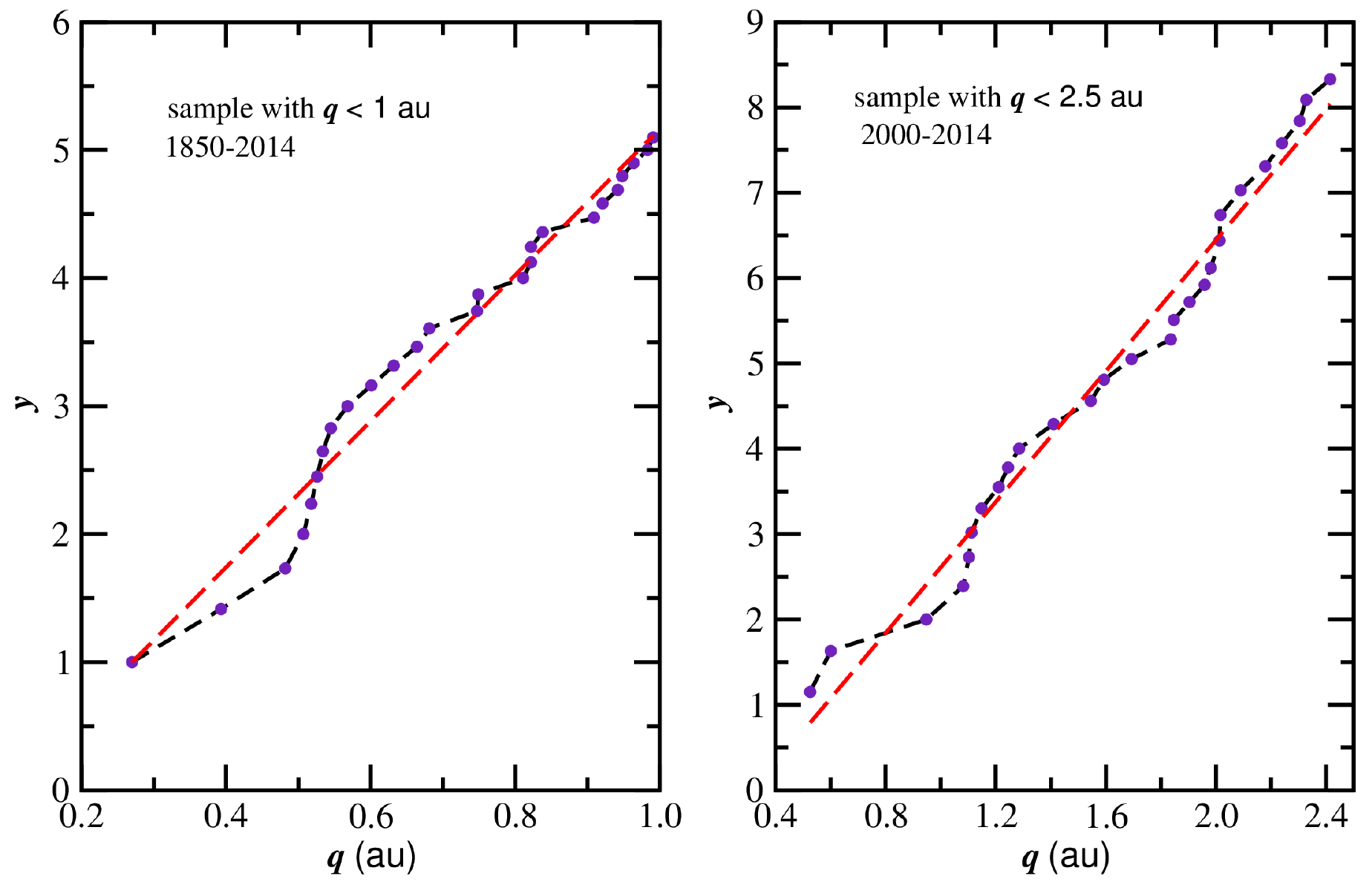}}
\caption{Linear fits for the square root of the cumulative number of comets with perihelion distances $<q$, $y=N_c^{1/2}(<q)$ as a function of $q$ for the sample of old LPCs with $q < 1$ au discovered during 1850-2014 (left-hand panel), and the corresponding sample with $q < 2.5$ au discovered during 2000-2014 (right-hand panel). The samples have been corrected for missed detections as explained in the text.}
\label{qcum_fit}
\end{figure}

We have tested linear fits to the samples of old LPCs described before that are assumed to be near complete. We have therefore considered the following samples: old LPCs observed during 1850-2014 with $q < 1$ au; old LPCs observed during 2000-2014 with $q < 2.5$ au; and the same as before but restricted to $q < 1.3$ au. The best linear fits are found by taking the square root of the cumulative number of comets with perihelion distances $<q$, $y=N_c^{1/2}(<q)$ instead of $N_c(<q)$. The linear fits for the samples of old LPCs with $q < 1$ au discovered during 1850-2014, and those with $q < 2.5$ au discovered during 2000-2014, are shown in Fig. \ref{qcum_fit}. The 1850-2014 sample has been corrected by multiplying it by a factor 1.4 that takes into account some missing comets, while the 2000-2014 sample has a correction factor that assumes that the probability of discovery of old LPCs, $p_D$,  falls linearly between $q=0$ and $q=3$ au, being $p_D=1$ for $q=0$ au and $p_D=0$ for $q=3$ au. From the different tests, we find as the best linear fit
   
\begin{equation}\label{qcum}
y = a_1q - a_2
\end{equation}
where the best-fitted values for the coefficients are: $a_1 = 0.66 \pm 0.02$, and $a_2 = 0.11 \pm 0.04$. These coefficients are normalized such that the computed $N_c(<q)$ from eq.(\ref{qcum}) is expressed in units of comets yr$^{-1}$ that reach perihelion distances $<q$. Note that the linear fit is with $N_c^{1/2}$ and not with $N_c$, and that it is valid for $q > a_2/a_1 \simeq 0.167$ au. This condition takes into account the observed fact that no old comets are observed coming very close to the Sun due to their very short physical lifetime. The correlation coefficient between $y$ and $q$ was found to be above 0.94 for all the considered samples. We will next assume that eq.(\ref{qcum}) is valid up to $q=6$ au, so we can extrapolate it to get the population of old LPCs reaching Jupiter's region.\\

From eq.(\ref{qcum}) we find: $N_c(<1.3) = 0.56$ comets yr$^{-1}$, $\Delta N_c (1.3-2.5) = N_c(<2.5) - N_c(<1.3) = 1.81$ comets yr$^{-1}$,  $\Delta N_c (4-6) = N_c(<6) - N_c(<4) = 8.42$ comets yr$^{-1}$. We thus have the following proportions among the three populations: $N_c (<1.3)$ : $\Delta N_c (1.3-2.5)$ : $\Delta N_c (4-6)$ $\simeq$ 1 : 3.25 : 15.

\subsection{The integration method}

We have integrated the massless test bodies under the gravitational potential of the Sun and the planets from Venus to Neptune for 4 Myr in the case of the populations with $q<2.5$ au, and for 10 Myr for the sample with $4 < q < 6$ au using the code EVORB \citep{Fern02}. The mass of Mercury was added to the Sun. We assumed that a ``collision'' with the Sun occurred when the comet reached a perihelion distance smaller than the Roche's limit 

\begin{equation}\label{rtide}
r_{tide} \simeq 2.5\left(\frac{\rho_{\odot}}{\rho_c}\right)^{1/3} R_{\odot} \simeq 0.0173 \mbox{ au,}
\end{equation}
where $\rho_{\odot}$ and $\rho_c$ are the bulk densities of the Sun and the comet, and $R_{\odot}$ is the Sun's radius. We assume for the comet nucleus a bulk density $\rho_c = 0.4$ g cm$^{-3}$ \citep{Sosa09}. When the test body collides with the Sun, or one of the planets, or reaches  $r > 1000$ au is eliminated from the integration. In every perihelion passage an output with the body's orbital elements is generated in order to follow its orbital evolution through successive perihelion passages. \\

We then generated three files for the purely dynamical evolution of the three samples, considering only gravitational effects, assuming that there are not other losses than ejections and collisions. We consider first the orbital energy. The energy distribution histogram of each sample is constructed taken the {\it residence time} as a proxy. The residence time is computed as follows: when the comet reaches perihelion ($< 1.3$ au) if its osculating energy $x$ falls within the range $x_i< x<x_i+\Delta x$, the residence time will be its orbital period $=x^{-3/2}$, assuming that $x$ does not change during the comet's revolution (this is in statistical terms a reasonable approximation). Note that the residence time within an energy range $(x_i, x_i+\Delta x)$ is expressed in years, and can be understood as the likelihood that a comet will be observed at a certain time within this energy range. The energy distribution histogram of the considered sample will be then equivalent to the distribution of the sum of the residence times of all the sample comets within energy ranges $(x_i, x_i+\Delta x)$.\\

In order to obtain the total contribution of the three samples to the near-Earth population ($q < 1.3$ au), we combined the three energy-histograms from the samples S1, S2 and S3 weighting each one of them with a factor that takes into account the estimated proportions of the real populations, namely, sample S1 ($q<1.3$ au) : sample S2 ($1.3<q<2$ au) : sample S3 ($4<q<6$ au) $\rightarrow$ 1 : 3.25 : 15.\\

Following a similar procedure, we built the inclination and perihelion distance distributions of the samples S1, S2 and S3 and their combined histograms.

\subsection{Physical evolution}

We considered the sublimation lifetime in terms of the number of perihelion passages before its physical disintegration $N_{phys}$. From thermophysical models we can derive the sublimation rate and the mass loss $\Delta m_P$ per orbital revolution of a water-dominated comet nucleus as done, for instance, by \citet{Disi09}. These authors could fit a polynomial expression for $\Delta m_P$ as a function of $q$. Since the erosion rate per orbital revolution is $\Delta h_P \sim \Delta m_P / \rho_c$, the physical lifetime of the comet of radius $R$ should be of the order $N_{phys} \sim R/\Delta h_P$ and can also be fitted to a polynomial of $q$. The polynomial fit, normalized to $N_{phys} = 100$ for $q=1$ au, is found to be

\begin{equation}\label{nphys}
N_{phys}(q) = 83.92q^6 - 493.2q^5 + 1122.7q^4 - 1163.3q^3 + 550.7q^2 -11.96q + 11.23
\end{equation}
We can normalize to other values of $N_{phys}$ just simply multiplying eq.(\ref{nphys}) by a scale factor.\\

We performed the simulation of the physical evolution as follows. From the file containing the whole pure dynamical evolution of each sample, as described in the previous section, we read the number of the perihelion passage of each test body and its orbital elements. We apply physical losses only if $q<2$ au using eq.(\ref{nphys}) by a Monte Carlo method. The procedure was the following: if a test body fulfilled the condition that $q<2$ au at a certain time, we then took a random number $z$ in the interval (0,1) and, if $z < 1/N_{phys}(q)$, we assumed that the body was physically destroyed and eliminated from the sample, otherwise we kept it in the simulation. If a test body survived the passage and have in addition $q<1.3$ au, we recorded its residence time with energy $x$ as explained above. Then we combined the energy distributions of the three histograms from the samples S1, S2 and S3 weighted by the factors 1, 3.25 and 15, as done before for the purely dynamical case. We thus obtained the energy distribution of the total population of surviving bodies with $q<1.3$ au for different physical lifetimes. We have run the simulations with physical losses for different values of $N_{phys}(q=1): 50, 100, 200, 500, 1000$ revolutions.\\

In a similar way we computed the distributions of $q$ and $i$.\\

We repeated the numerical simulations for the entirely fictitious samples F1, F2 and F3, defined in Section 3.1, and computed the distributions of $x$, $i$ and $q$ for the purely dynamical case and for the different physical lifetimes $N_{phys}(q=1)$ quoted before.

\section{The results}

\subsection{End states and dynamical paths}
\label{ends}

The dynamical half time of the samples S1, S2 and S3 are found to be:  0.37 Myr, 0.58 Myr and 0.75 Myr respectively, considering only the purely dynamical evolution. From the initial number of 1000 test bodies for each sample, the number of survivors at the end of the integration were 14, 32 and 5 for the samples S1, S2 and S3 respectively. Most of the objects are eliminated in hyperbolic orbits but a small percentage ($\simeq 4.7\%$) are eliminated by collision with the Sun. Collisions with the planets are very rare.\\

\begin{figure}[h]
\resizebox{12cm}{!}{\includegraphics{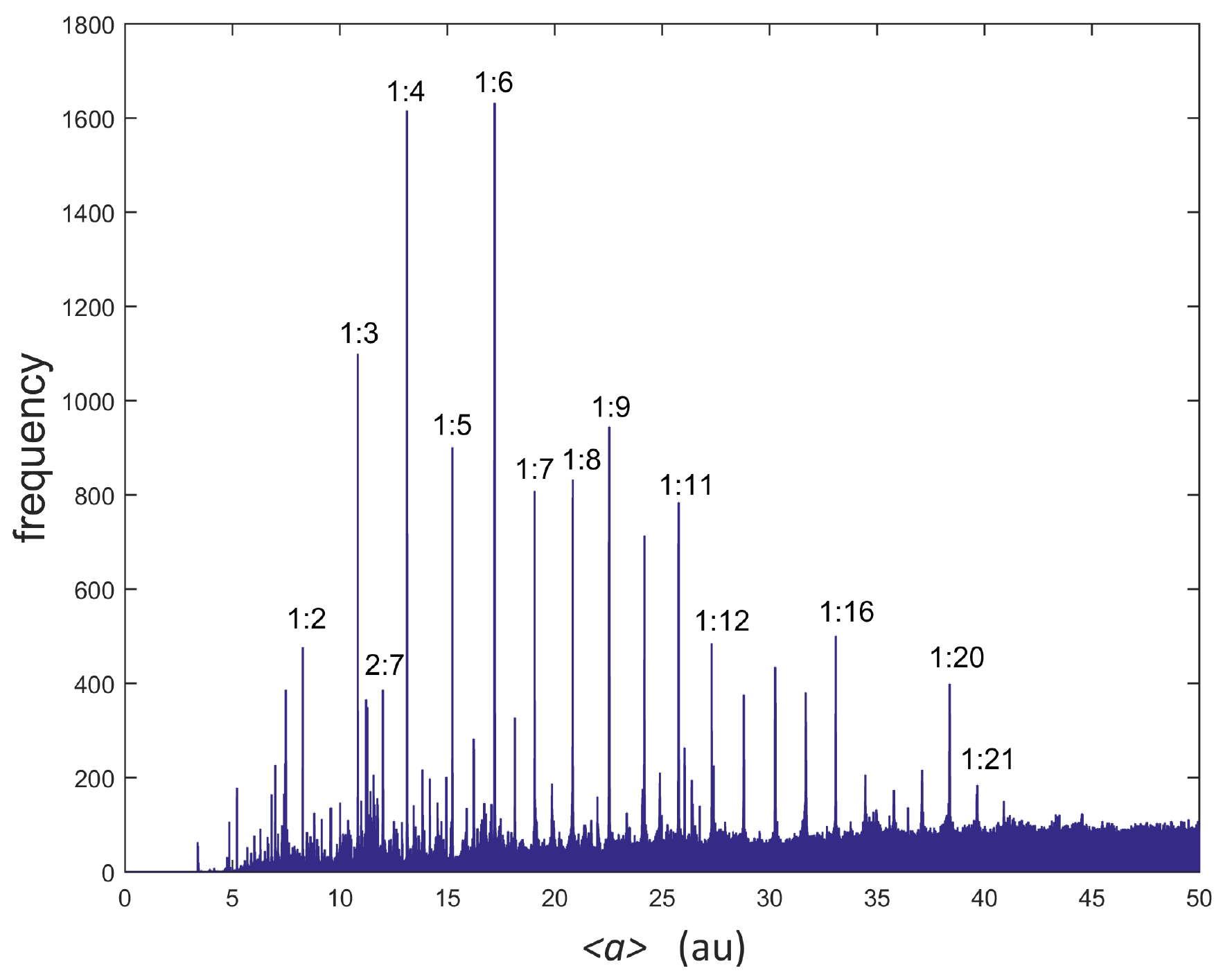}}
\caption{Frequency distribution of the time-average $<a>$ (over $10^4$ yr) of comets reaching or staying with $q < 1.3$ au, for a sample of LPCs with initial $q<2.5$ au.}
\label{resonances}
\end{figure}

A fraction of the comets evolve into more tightly bound orbits where the systematic gravitational effects of the planets become more relevant, in particular, trapping in mean motion resonances (MMRs). Resonances are easily detected because the time evolution of the semimajor axis changes its behavior from erratic to a small amplitude oscillation around a constant mean value $<a>$. In order to study the relevance of the MMRs, we calculate the mean value of the orbital elements in time intervals of $10^4$ yr using all orbital states from the numerical integrations. We then construct histograms of $<a>$ with bins of 0.01 au. These histograms show a distribution of $<a>$ strongly influenced by captures in MMRs with the giant planets, mainly Jupiter. In Fig. \ref{resonances} we present a histogram collecting all the $<a>$ values from the numerical integrations of a sample of 1000 objects with $q < 2.5$ au comprising the observed old LPCs described before (cf. Section 3.1) cloned in order to account for the weighting factor $P/164$ and the proportion 1:3.25 between the populations with $q < 1.3$ au and $1.3 < q < 2.5$ au. The most prominent feature is a series of spikes corresponding to captures mostly in exterior resonances 1:N and 2:N with Jupiter which are the strongest ones. The general trend of the spikes for semimajor axes greater than 12 ua is in good agreement with strengths of the resonances in that region as can be deduced from \citet{Gall06}. By comparing the frequency of $<a>$ in the spikes with that in the background (the bottom black region of Fig. \ref{resonances}), we find that at least $12\%$ of the orbital evolution of comets reaching semimajor axes $a < 50$ au is done inside two-body exterior MMRs with Jupiter.\\ 

\begin{figure}[h]
\resizebox{12cm}{!}{\includegraphics{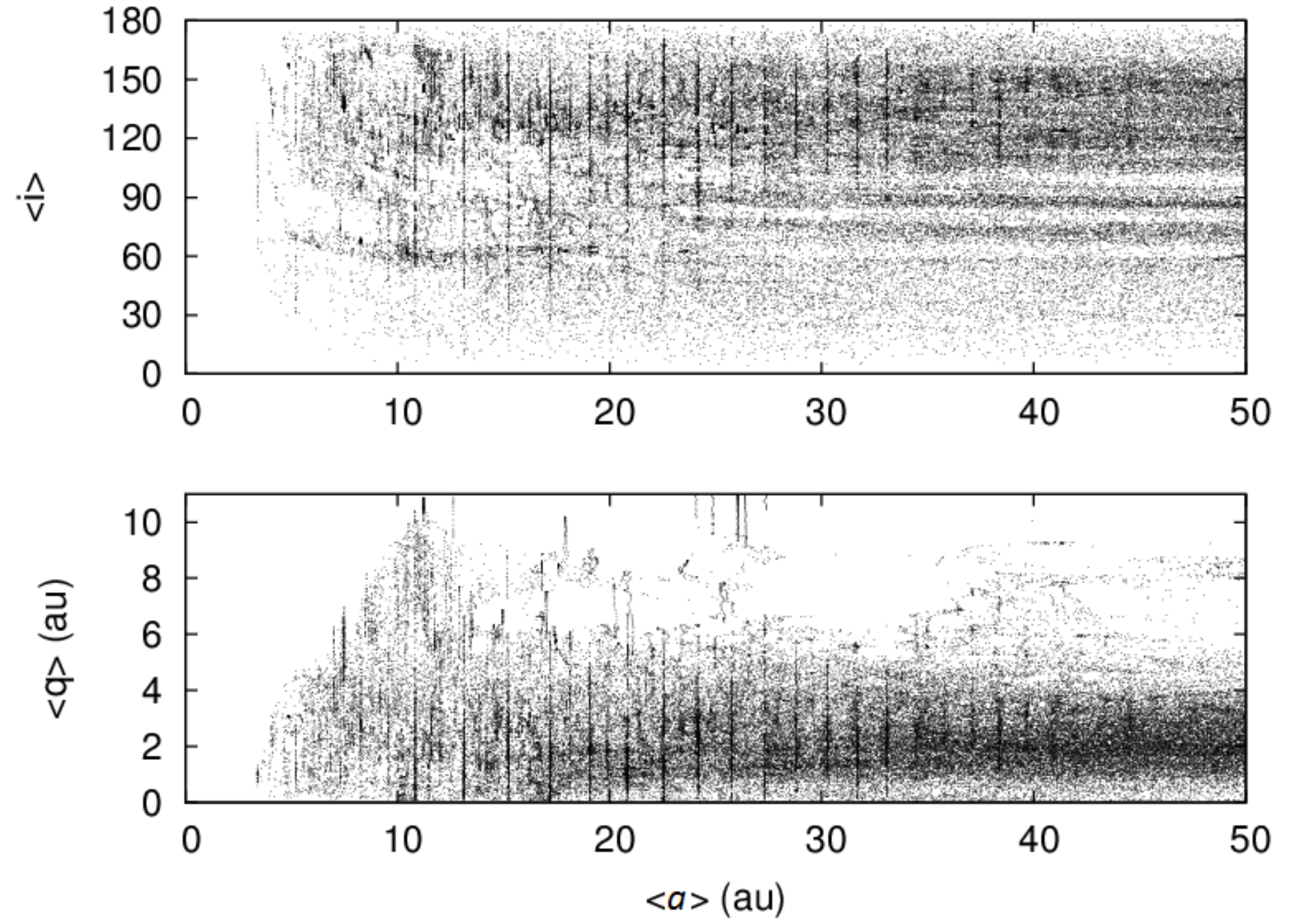}}
\caption{Average inclination (top panel) and average perihelion distance (bottom panel) versus average semimajor axis, taken within interval of $10^4$ yr, for a sample of 1000 test bodies consisting of the observed LPCs with initial $q<2.5$ au and their clones. Several MMRs are prominent in particular for comets in retrograde orbits.}
\label{mediosia}
\end{figure}

Most of the objects captured in MMRs are evolving in retrograde orbits, a fact that is consistent with the results showed by \citet{NM15} (Fig. \ref{mediosia}). The results obtained here are also consistent with the study of the restricted three-body problem Sun-Jupiter-comet by \citet{Cham97} in which he found that stable librations can occur for up to the 1:9 MMR with Jupiter ($P \simeq 107$ yr) for direct low-inclination comets, while stable librations can extend to the 1:21 MMR ($P \simeq 250$ yr) for retrograde orbits. This can be understood in terms of the greater or lesser gravitational perturbing influence of Jupiter for direct or retrograde orbits \citep{Fern05}.

\subsection{Sungrazing states}

The Kozai mechanism is the main responsible for large amplitude oscillations in $q$ and $i$, specially for particles also captured in MMRs. This is clearly showed in Fig. \ref{mediosia} where large excursions in $q,i$ occurs while  $<a>$  is locked in a MMR. This mechanism is responsible for driving many old LPCs and HTCs toward the Sun, as was originally discussed in depth by \citet{Bail92}.\\

\begin{figure}[h]
\resizebox{10cm}{!}{\includegraphics{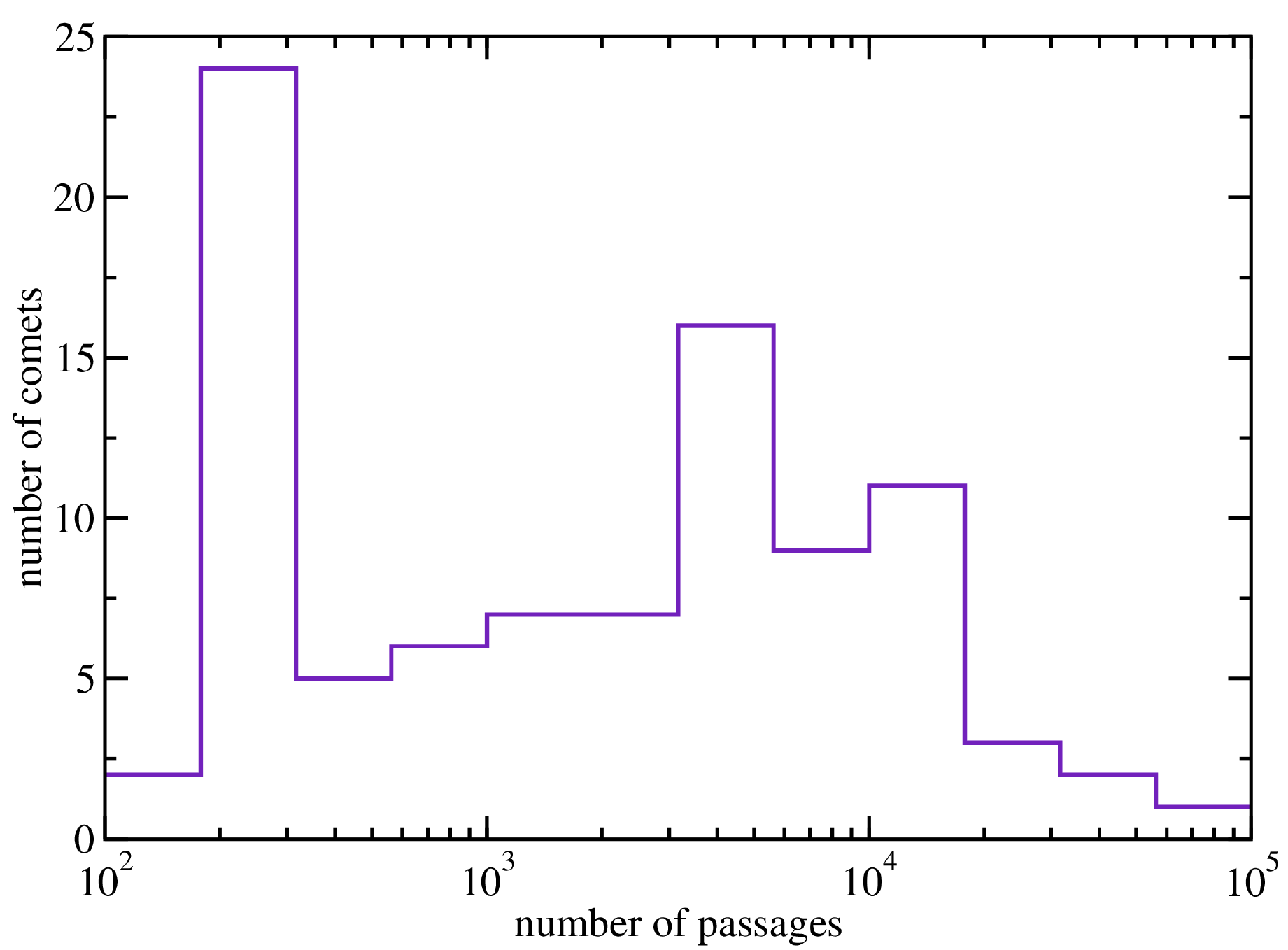}}
\caption{Number of perihelion passages required for the comets of the sample S1 to reach a perihelion distance for ``collision'' with the Sun (the Roche's radius $\sim 3 .7\mbox{R}_{\odot} \simeq 0.0173$ au) from a perihelion distance $q = 0.2$ au.}
\label{sungrazer}
\end{figure}

As mentioned before, about 4.7\% of our computed LPCs end up their evolution colliding with the Sun in the purely dynamical case. It is noteworthy to mention that a comet ``collides'' with the Sun (i.e. reaches a perihelion distance $q < 0.0173$ au) after a smooth decrease in its perihelion distance. We find that the average number of perihelion passages required to decrease the perihelion distance from 0.2 au to 0.0173 au is typically from several hundreds to thousands revolutions (Fig. \ref{sungrazer}). In all the studied cases the approach to the Sun took more than one hundred revolutions.

\subsection{Distribution of the energy}
\label{ener}

\begin{figure}[h]
\resizebox{12cm}{!}{\includegraphics{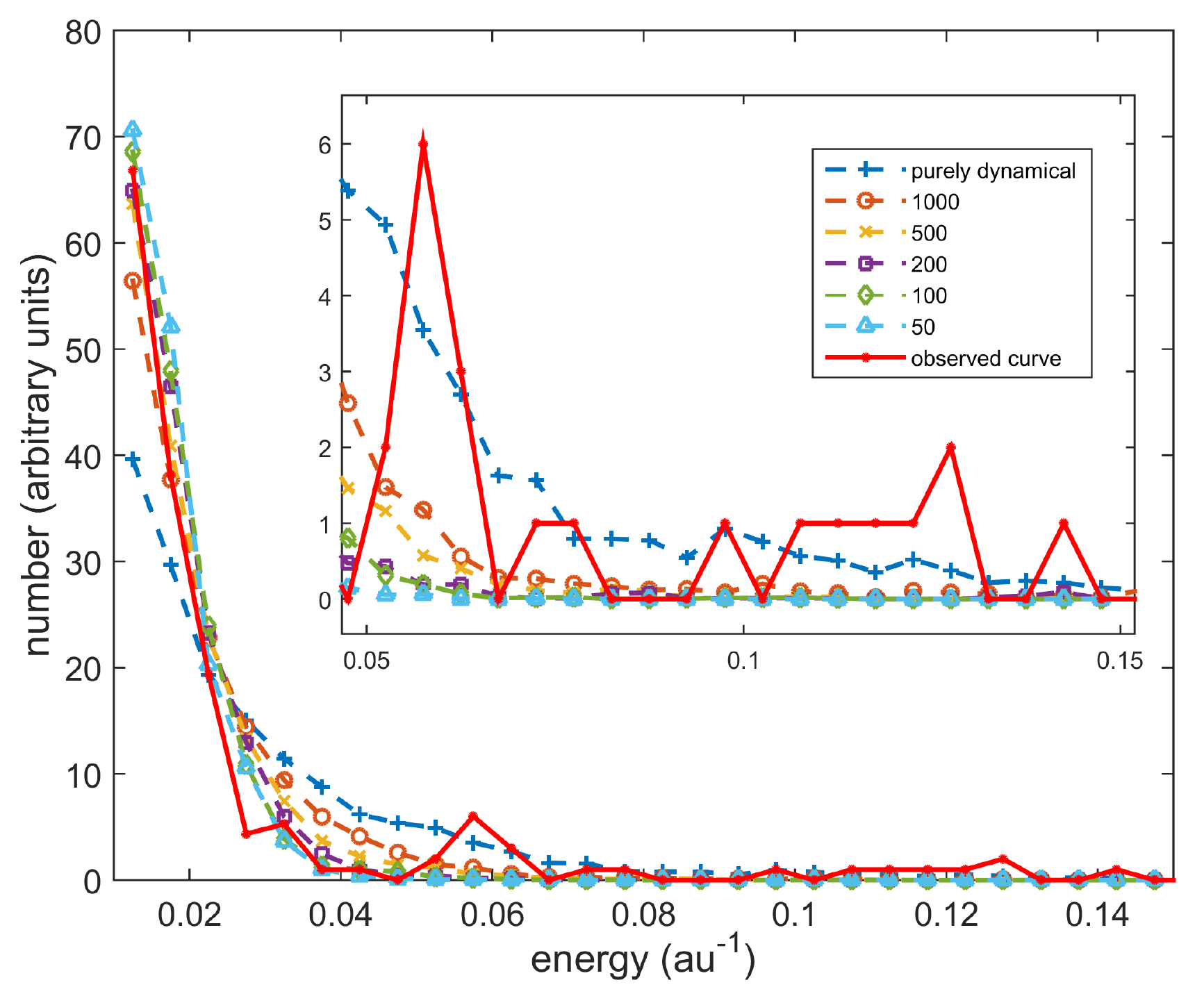}}
\caption{Energy distribution of the computed LPCs (weighted sum of samples S1, S2 and S3) that reach or stay with $q < 1.3$ au for a purely dynamical evolution and for different physical lifetimes $N_{phys}(q=1)$ (dashed curves), as indicated in the inset panel. The observed energy distribution is represented by the solid curve.}
\label{energy}
\end{figure}

In Fig. \ref{energy} we show the computed energy distribution of comets that reach or stay with $q < 1.3$ au, obtained following the procedure described in the previous section for a purely dynamical evolution and for physical lifetimes $N_{phys}(q=1) =$ 50, 100, 200, 500 and 1000. The computed energy distributions are overlapped with the observed one (solid curve) as given by Fig. \ref{energies}. We tried to assess in which case we obtain the best match between the computed and the observed distribution by considering the minimum of the function: $\Delta^2(O-C) = \sum^{0.15}_{0.01} [N_o(x_i) - N_c(x_i)]^2$. where $N_o(x_i)$ and $N_c(x_i)$ are the observed number of comets and the computed residence times (as a proxy of the number of comets) in the energy range $x_i \pm \Delta x/2$, with the width $\Delta x = 0.01$ au$^{-1}$.\\

By inspecting Fig. \ref{energy} we find that the match between the observed and the computed energy distributions is quite poor, either for the purely dynamical case, as well as for the cases with different physical lifetimes. Only the purely dynamical evolution produces a good number of comets with $x \gsim 0.04$ au$^{-1}$ (see inset panel in Fig. \ref{energy}), but we find a total mismatch in the range $0.01 < x \lsim 0.04$ au$^{-1}$ where the computed distribution appears as too flattened as compared to the observed one. For finite physical lifetimes we improve the fitting in the range $0.01 < x \lsim 0.04$ au$^{-1}$, but for energies $x \gsim 0.04$ au$^{-1}$ we find too few computed comets as compared with the observed ones. In particular, it is not possible to reproduce from our simulations the clustering of comets observed at energies about $0.055-0.065$ au$^{-1}$. We can get a rather good match between the observed and the computed energy distribution only within the left narrow portion $0.01 < x \lsim 0.04$ au$^{-1}$. Within this range the best matches are found for physical lifetimes $N_{phys}(q=1) \sim 100-500$ revolutions.

\subsection{Distribution of the inclinations}
\label{incl}

\begin{figure}[h]
\resizebox{12cm}{!}{\includegraphics{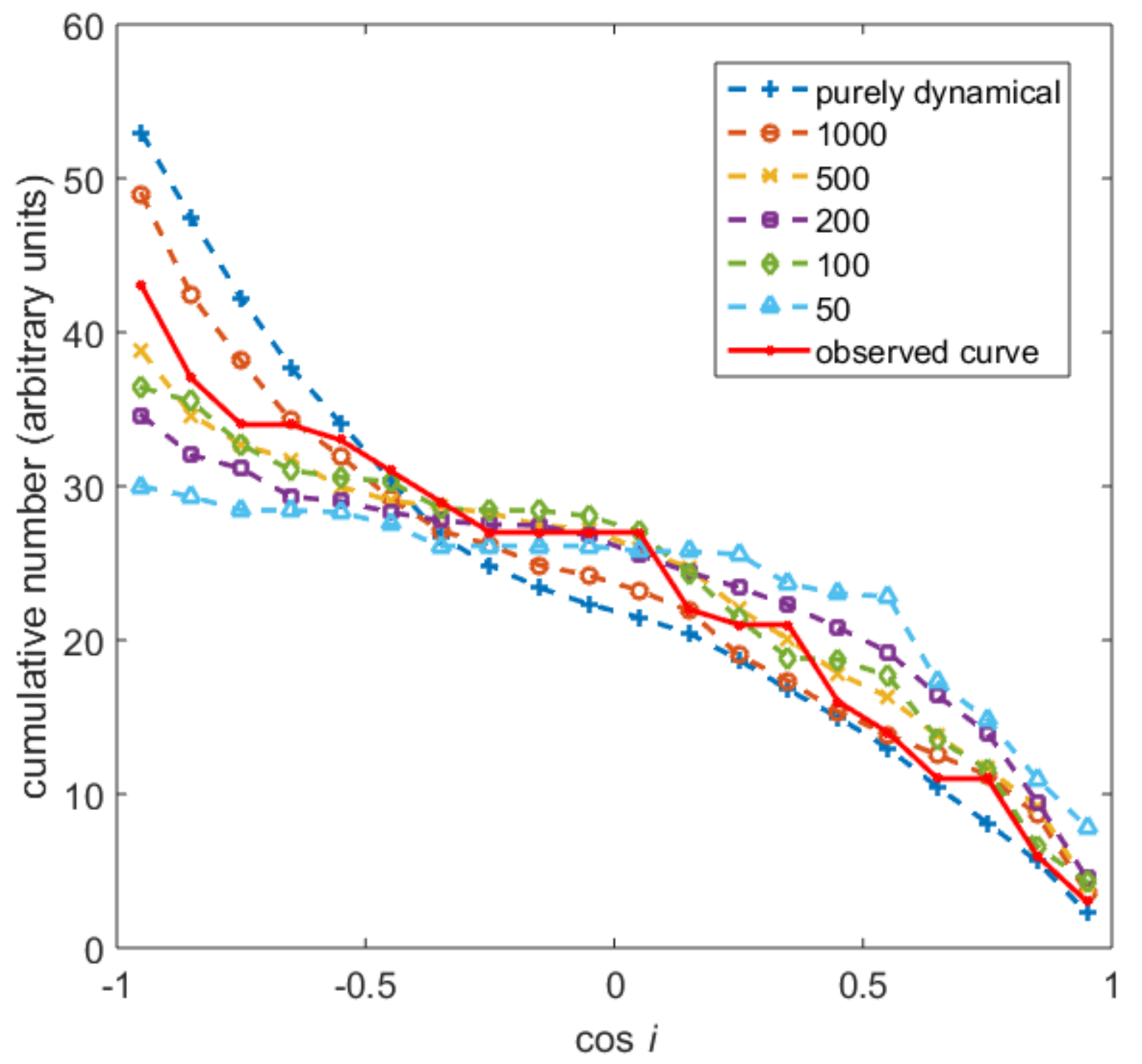}}
\caption{Cumulative distribution in $\cos{i}$ (going from $\cos{i}=+1$ to $\cos{i}=-1$) of the computed comets (weighted sum of samples S1, S2 and S3) that reach or stay with $q < 1.3$ au and whose orbits stay of long-period ($200 < P < 1000$ yr) for a purely dynamical evolution and for different physical lifetimes, $N_{phys}(q=1)$, indicated in the inset panel (dashed curves). The cumulative distribution of the observed old LPCs is represented by the solid curve.}
\label{incli}
\end{figure}

In Fig. \ref{incli} we plot the cumulative number of comets with values $> \cos{i}$, starting from $\cos{i}=1$ to $\cos{i}=-1$, for the purely dynamical case (solid curve) and for the physical lifetimes quoted above (dashed curves). As explained before, we take as a proxy of the number of comets having values $> \cos{i}$ the sum of the residence times of all the test comets of samples S1, S2, S3 having values $> \cos{i}$ and also fulfilling the conditions: $q < 1.3$ au and $200 < P < 1000$ yr. We see clear departures from a random distribution of $\cos{i}$, consisting in a flattening of the curves toward retrograde orbits, either for the observed curve, or for the computed ones for physical lifetimes $N_{phys}(q=1) \lsim 500$ revolutions. The differences (O-C) are more subtle. For $N_{phys}(q=1) \gsim 1000$ revolutions the computed model produces some excess of retrograde comets as compared to the observed one, that show up as a fast increase in the curves as we approach $\cos{i}=-1$. Again, the best fittings are found for physical lifetimes $N_{phys}(q=1) \sim 100-500$ revolutions.

\subsection{Distribution of the perihelion distances}
\label{perih}

The distribution of perihelion distances of our computed LPCs that reach $q < 1.3$ au is shown in Fig. \ref{perih}. Again, we compare the computed cumulative $q$-distributions (dashed curves) with the observed one (solid curve). The match is very poor for the purely dynamical evolution since it produces too many comets with $q \lsim 0.4$ au and too few for $q \gsim 0.8$ au. If we compare the (O-C) distributions for the different physical lifetimes, the best fit is obtained for $N_{phys}(q=1) \sim 500$ revolutions.\\

\begin{figure}[h]
\resizebox{12cm}{!}{\includegraphics{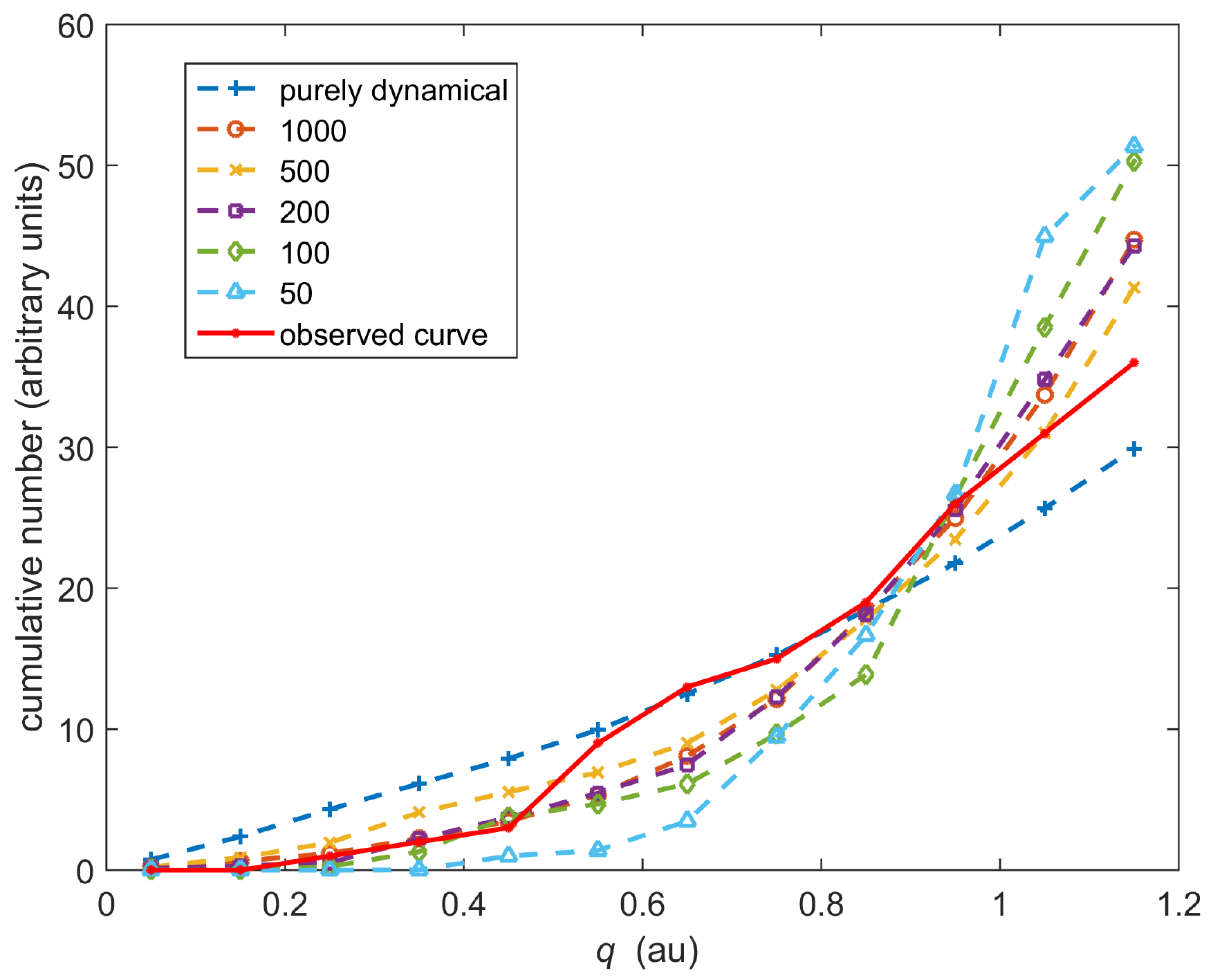}}
\caption{Cumulative distribution of the perihelion distances of the computed comets (weighted sum of samples S1, S2 and S3) whose orbits stay of long-period ($200 < P < 1000$ yr) for a purely dynamical evolution and for different physical lifetimes, $N_{phys}(q=1)$, indicated at the upper left corner (dashed curves). The cumulative $q$-distribution of the observed old LPCs is represented by the solid curve.}
\label{perih}
\end{figure}

Summing up, if we combine the different fittings for $x$, $i$ and $q$, we find some dispersion, though the different results for $N_{phys}(q=1)$ tend to cluster around a few hundred revolutions, that we can narrow down to $\sim 200-300$ revolutions as the best compromise.

\subsection{Evolution with nongravitational forces}
\label{ngf}

As described in Section 3.1, we performed some runs with the samples S1, S2 and S3, adding to the equations of motion nongravitational terms. In statistical terms, the results did not show any significant difference, in particular with what was our main concern, namely the computed distributions of $x$, $i$ and $q$. These tests reassured us that nongravitational forces can be disregarded without affecting appreciably the validity of our conclusions. 

\subsection{The Tisserand parameter versus the energy}
\label{Tissvsx}

\begin{figure}[h]
\resizebox{15cm}{!}{\includegraphics{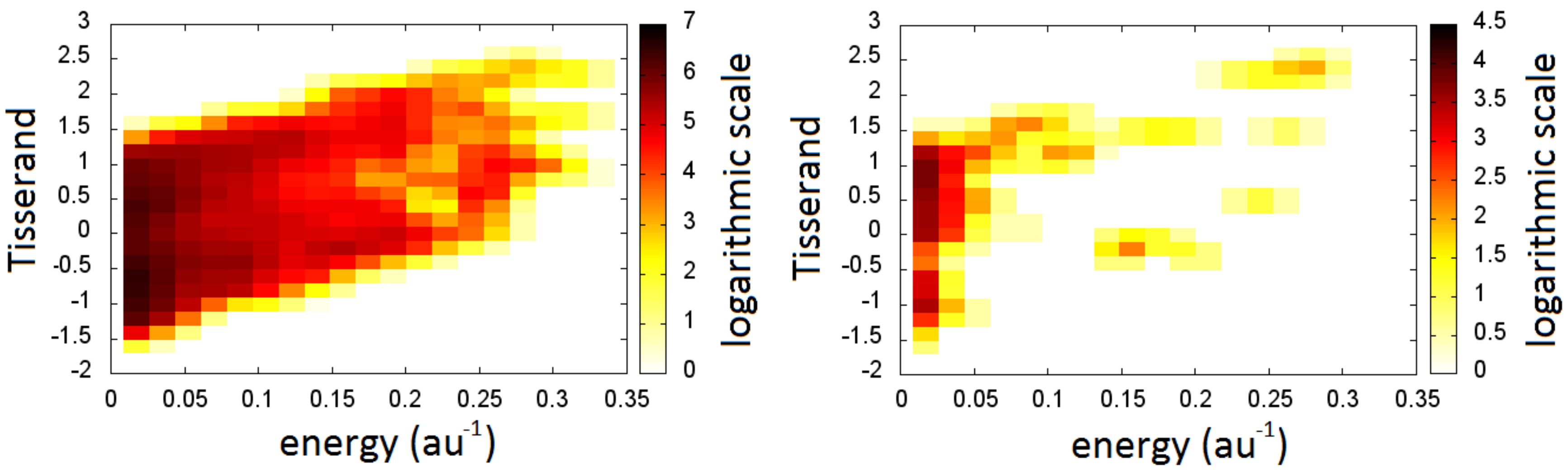}}
\caption{Distribution of residence times in the plane orbital energy versus Tisserand parameter of the computed LPCs (weighted sum of samples S1, S2 and S3) evolving within $q < 1.3$ au. These cases correspond to a purely dynamical evolution (left-hand panel), and to a physical lifetime of $N_{phys}(q=1) = 200$ revolutions (right-hand panel).}
\label{tvsxlpc}
\end{figure}

As expected, the Tisserand parameter with respect to Jupiter, $T_J$, of LPCs with $q < 1.3$ au tend to keep below the threshold $T_J = 2$ that roughly divides the HTC population from the JFC population (Fig. \ref{tvsxlpc}). If we consider the case of a purely dynamical evolution, only 3.3\% of the comets that reach energies $x > 0.136$ au$^{-1}$ (orbital periods $P < 20$ yr) acquire Tisserand parameters $T_J > 2$, and thus they could be classified as JFCs. If we apply a physical lifetime $N_{phys}(q=1) = 200$ revolutions, the percentage raises to 13.4\%, though we note that only 2.1\% of the computed comets reach energies $x > 0.136$ au$^{-1}$. We conclude that when we consider a finite physical lifetime the probability of a LPC to reach a JFC orbit is extremely low ($<<1\%$).

\subsection{Comparison of the previous results with those obtained with the purely fictitious samples of LPCs}
\label{FictLPCs}

We recall that the previous results were obtained from the samples S1, S2, S3, in which S1, S2 are based on the observed samples of old LPCs with $q < 2.5$ au. Since the observed samples might have some uncorrected bias, we wanted to check the previous results with random fictitious samples as defined in Section 3.1 (samples F1, F2 and F3). The computed results obtained with the latter samples were consistent with those shown in Figs. \ref{energy}, \ref{incli} and \ref{perih} with some slight differences that we explain below.\\

The series of energy-distribution histograms for the purely dynamical evolution and for physical lifetimes $N_{phys}(q=1) =$ 50, 100, 200, 500 and 1000, follow the same pattern as Fig. \ref{energy}, but the best fitting was obtained for $N_{phys}(q=1) \simeq 500-1000$, i.e. somewhat longer than that obtained before. This can be explained since the fictitious comets all start with an initial $P=10^3$ yr, while the observed samples and their clones with $q < 2.5$ au (samples S1 and S2) start in a more evolved stage with orbital periods ($200 < P < 10^3$ yr), so their physical lifetimes should be somewhat shorter.\\

As regards the inclination distribution, the best fitting is for $N_{phys}(q=1) = 500$ revolutions. The only stricking feature is a discrepancy in the number of comets with inclinations close to $180^{\circ}$: we note some excess in the observed sample of LPCs that it is not reproduced with the computed random samples. We will come back to this point below. In agreement with the previous result, the best fitting for the $q$-distribution is also found for $N_{phys}(q=1) =500$ revolutions.\\

These new computer runs with fully fictitious samples confirm our previous conclusion as regards the imposibility to reconcile the steep slope in the energy distribution in the range $0.01 < x < 0.04$ au$^{-1}$ with the observed long tail of comets with energies $x > 0.04$ au$^{-1}$.

\section{The search for an alternate source for HTCs}
\label{source}

Our preliminary conclusion was that the old LPCs coming close to the Earth or Jupiter were not the immediate precursors of comets with energies $x \gsim 0.04$ au$^{-1}$ and Tisserand parameters $T < 2$ that we identified with the HTCs. Either we find a rather flat energy distribution with a long tail of larger binding energies ($x \gsim 0.04$ au$^{-1}$) when we consider a purely dynamical evolution, or we are able to obtain a steep decrease in $x$ in the range $0.01 < x < 0.04$ au$^{-1}$, consistent with observations, but too few comets with larger binding energies for finite physical lifetimes ($N_{phys}(q=1) \lsim 500$ revolutions). Facing this unexpected result, we tested other potential source regions for the observed HTCs. We then proceeded to evaluate if the Centaur population with perihelia close to Jupiter's orbit could be a more suitable source of HTCs. Centaurs are a transient population evolving from the trans-neptunian region to the inner planetary region. In particular, the Scattered Disk Objects (SDOs) may be the main source of Centaurs \citep{Dunc97, Disi07}, though it might be some significant contribution from Plutinos in the 2:3 MMR with Neptune \citep{Morb97}.\\

We chose a sample of 204 Centaurs with $2.5 < q < 8$ au, $a > 6$ au, and $Q < 30$ au, taken from {\it JPL Solar System Dynamics, ssd.jpl.nasa.gov}, and cloned them four times each, varying their orbital elements following the same procedure implemented for the clones of LPCs (cf. Section 3.1). In the selected sample of observed Centaurs we considered D/1993 F2-A (Shoemaker-Levy 9) as the single representative of the 21 observed fragments of the comet. We obtained a total of 1020 bodies, so we removed 20 clones at random in order to keep a sample of 1000 test bodies for our simulations. We integrated the sample of 1000 test bodies for 10 Myr. We note that some Centaurs are classified as comets, because they show activity; while others are classified as asteroids since they have so far been observed inactive. We do not know if they represent different populations, or if they are active and inactive objects from the same population. \citet{Jewi09} finds that the difference between active and inactive Centaurs may lie in that the former have on average smaller perihelion distances. The exposure to a larger solar radiation flux might trigger the transition of amorphous to crystalline ice that drives the activity. In this study we grouped them in the same population, and assume that active as well as inactive Centaurs come from the trans-neptunian region, though some interlopers from the asteroid belt, or the Oort cloud cannot be ruled out.\\

\begin{figure}[h]
\resizebox{10cm}{!}{\includegraphics{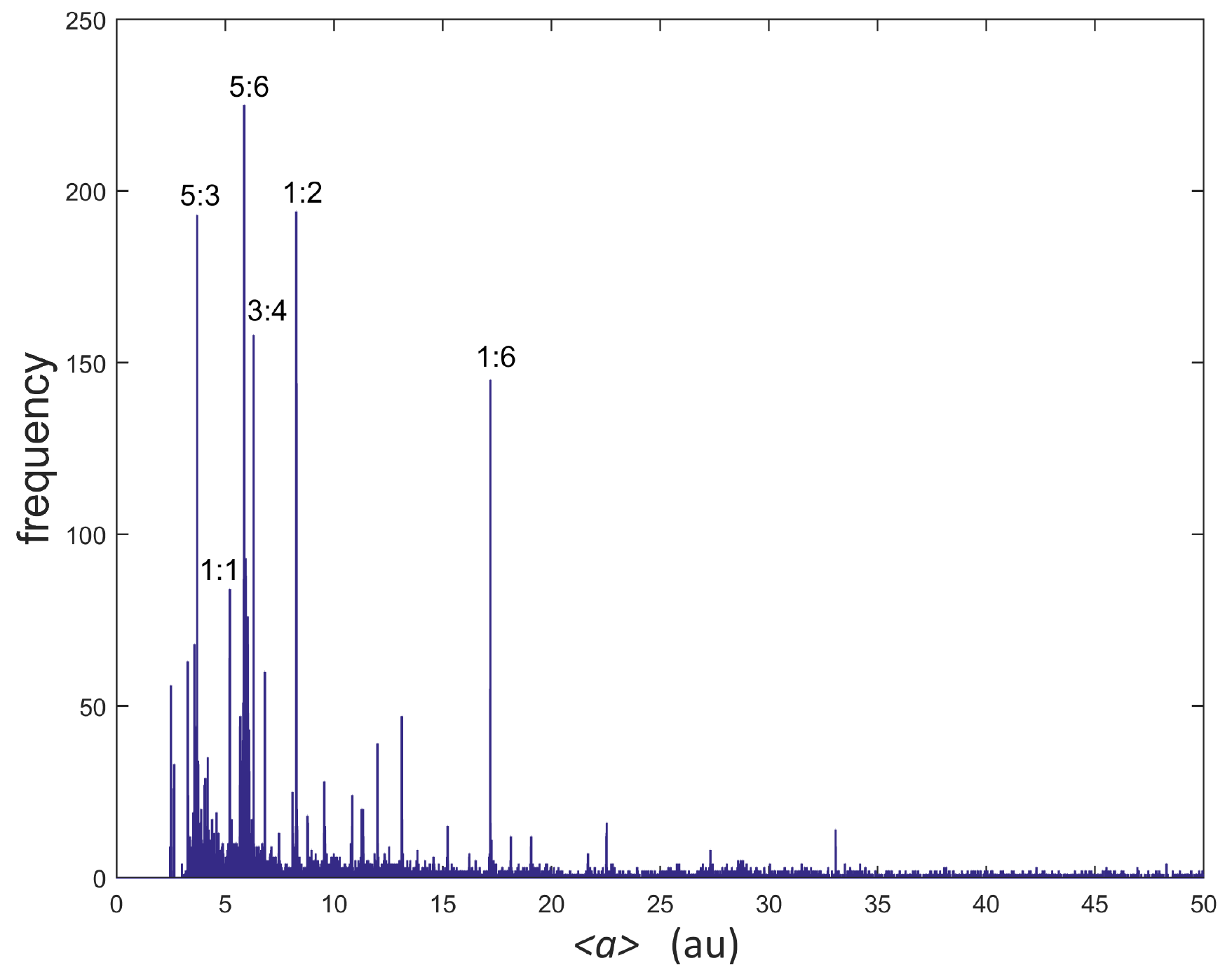}}
\caption{Frequency distribution of the time-average $<a>$ (over $10^4$ yr) of comets reaching or staying with $q < 1.3$ au, for the Centaur sample.}
\label{Res_Cent}
\end{figure}

We also considered the average $<a>$ within intervals of $10^4$ yr for the population of Centaurs and did the same analysis as that done for the LPCs. We found  that about $52.6\%$ of the orbital evolution of the Centaurs is done inside a two-body MMR. The dominance of some MMR, in particular the 5:3, 5:6, 3:4, 1:2, and 1:6, can be seen in Fig. \ref{Res_Cent}.\\

\begin{figure}[h]
\resizebox{14cm}{!}{\includegraphics{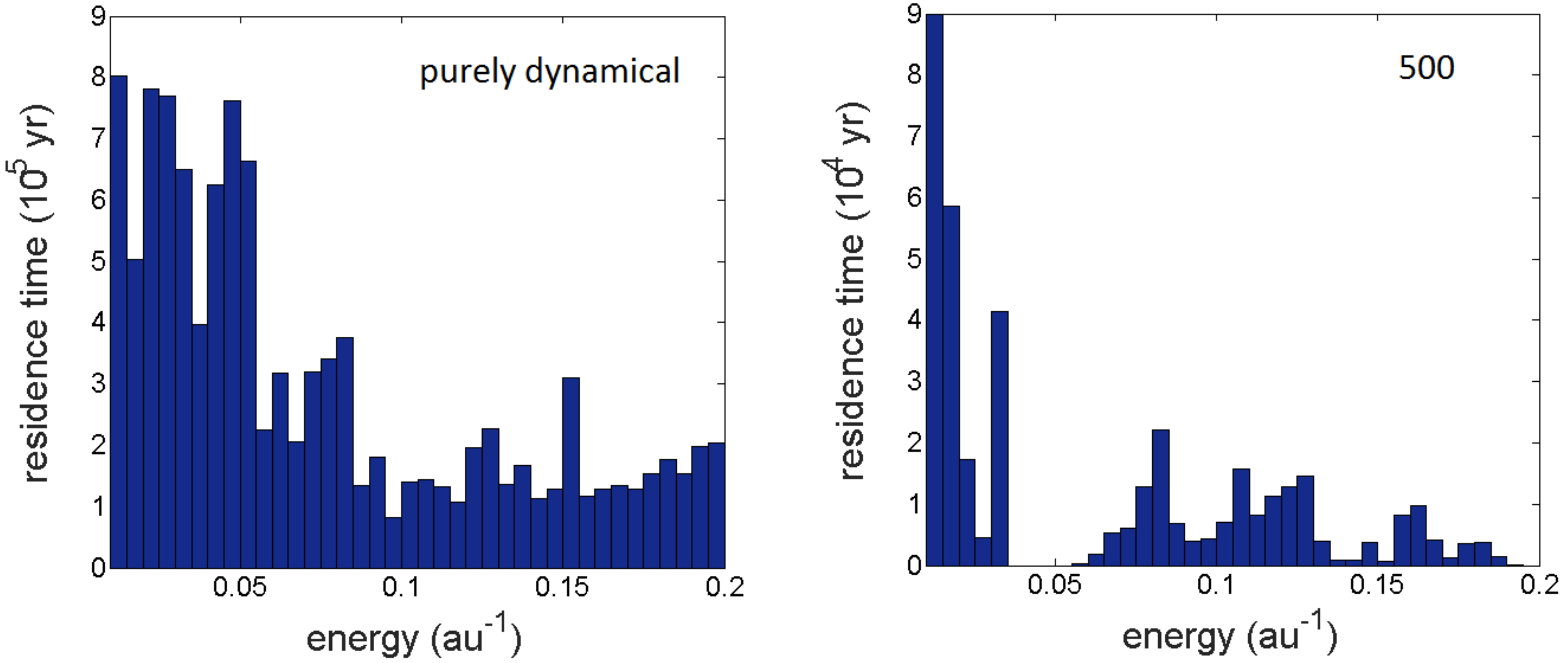}}
\caption{Energy distribution of the comets that reach perihelion distances $q < 1.3$ au from the Centaur sample for a purely dynamical evolution (left-hand panel) and for a physical evolution of $N_{phys}(q=1) = 500$ revolutions (right-hand panel).}
\label{xcent}
\end{figure}

\begin{figure}
\resizebox{12cm}{!}{\includegraphics{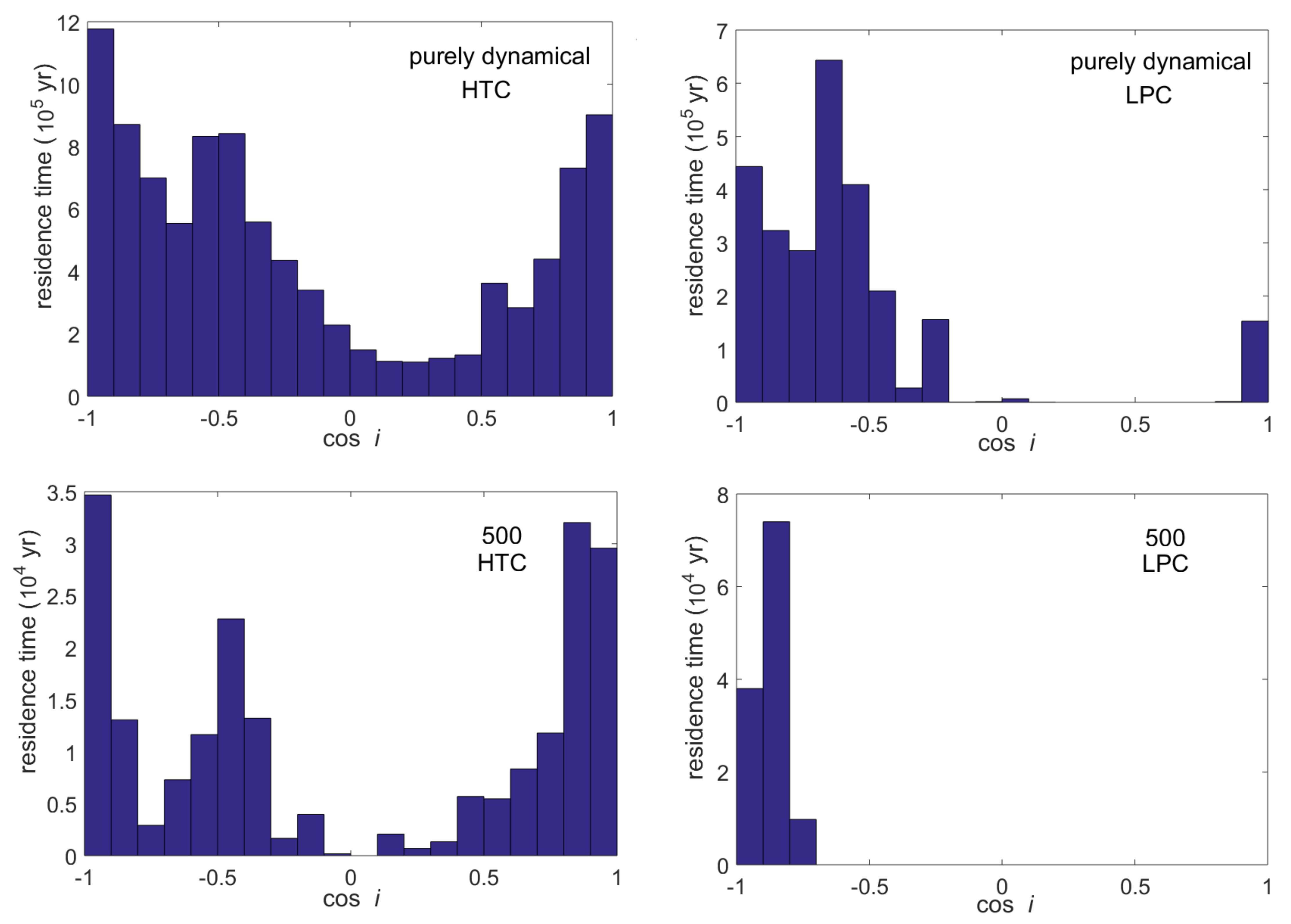}}
\caption{Inclination distribution of comets evolving from the Centaur sample that reach $q < 1.3$ au in Halley-type orbits (left-hand panels), and in long-period orbits (right-hand panels). We consider a purely dynamical evolution (upper row), and an evolution for a physical lifetime $N_{phys}(q=1) = 500$ revolutions (lower row).}
\label{icent}
\end{figure}

Fig. \ref{xcent} shows the energy distribution of Centaurs that reach $q < 1.3$ au as given by their residence times within given energy ranges (we remind that the number of observable comets is proportional to the overall residence time of all the sample comets within a given energy range). By contrast to the distribution observed in LPCs, the distribution of energies is more spread out in all the considered energy range. In particular, when we consider a finite physical lifetime ($N_{phys}(q=1) = 500$ revolutions), we observed that some clustering remains in the LPC energy range, other clustering is observed in the energy range $0.06 \lsim x \lsim 0.13$ au$^{-1}$ (orbital periods $21 \lsim P \lsim 70$ yr), i.e. typical of HTCs. Furthermore we find an extense tail of even larger binding energies that correspond to JFCs. As regards the inclination distribution, comets reaching HTC orbits (orbital periods $P < 200$ yr) show a clear predominance of low-inclination comets but with a long tail covering the rest of the inclination range, and a small grouping close to $180^{\circ}$ (Fig. \ref{icent}). On the other hand, the LPCs generated from the evolution of the Centaurs show a clear predominance of retrograde orbits.\\

\subsection{Tisserand parameter versus orbital energy}
\label{tiss}

\begin{figure}[h]
\resizebox{15cm}{!}{\includegraphics{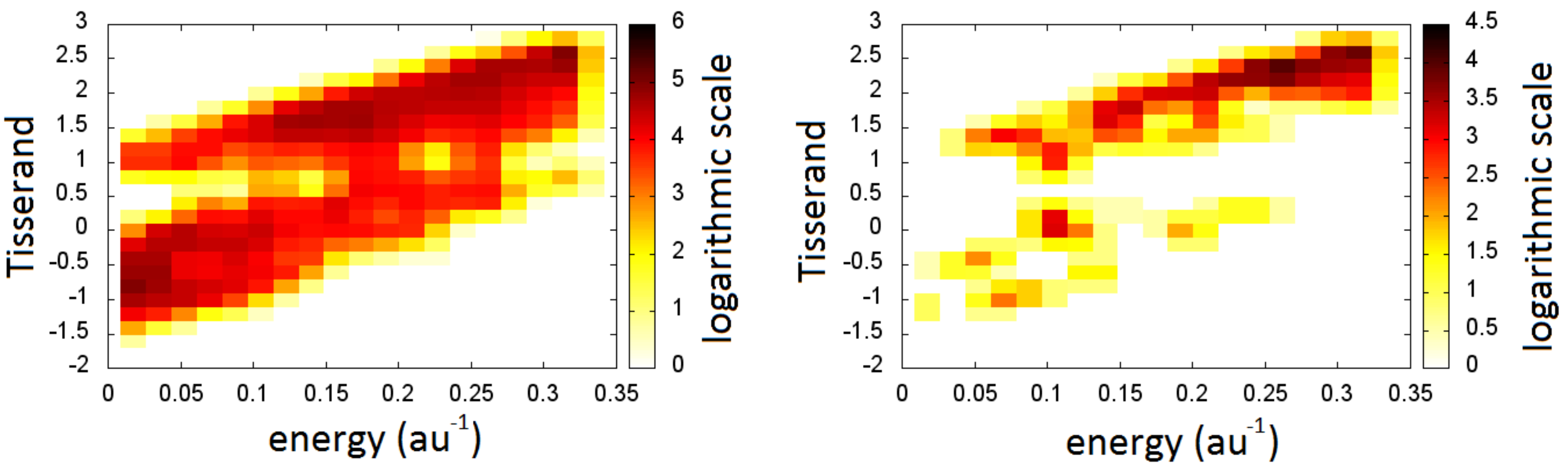}}
\caption{Distribution of residence times in the plane orbital energy versus Tisserand parameter of Centaurs evolving within $q < 1.3$ au. These cases correspond to a purely dynamical evolution (left-hand panel), and to a physical lifetime of $N_{phys}(q=1) = 500$ revolutions (right-hand panel).}
\label{tvsxcent}
\end{figure}

When we consider the dynamical evolution of Centaurs, we find that a substantial population (53.8\%) come to the region $q < 1.3$ au with energies $x > 0.136$ au$^{-1}$ ($P < 20$ yr). Among the Centaurs reaching $x > 0.136$ au$^{-1}$, 45.4\% have Tisserand parameters $T_J > 2$, i.e. they can be classified as JFCs, while the rest (54.6\%) have $T_J < 2$ and therefore are of the HTC type. These results are clearly in conflict with observations showing that most comets with $P < 20$ yr are of JFC type. Yet, these results are for a purely dynamical evolution. When we apply a finite physical lifetime, we observe significant changes. For instance, for $N_{phys}(q=1) = 500$ revolutions, among the Centaurs reaching $q < 1.3$ au the percentage with energies $x > 0.136$ au$^{-1}$ raises to 72.2\% and, among these, 83.9\% have Tisserand parameters $T_J > 2$, which is consistent with Centaurs being the immediate source of most JFCs. The evolution of the energy versus the Tisserand parameter of Centaurs for a purely dynamical evolution and for $N_{phys}(q=1) = 500$ revolutions is shown in Fig. \ref{tvsxcent}. On the other hand, for the energy range $0.0292 < x < 0.136$ au$^{-1}$ (periods $20 < P < 200$ yr) that we could identify with HTCs (always considering bodies that reach $q < 1.3$ au) all comets have Tisserand parameters $T_J < 2$.

\section{Discussion}

We could not obtain a good fit of our computed energy distribution of old LPCs and HTCs to the observed distribution. We could obtain good fits only within the narrow range $0.01 \lsim x \lsim 0.04$ au$^{-1}$ (or orbital periods $125 \lsim P \lsim 1000$ yr) with finite physical lifetimes ($\sim 200-300$ revolutions for a comet with a standard $q = 1$ au). Yet, with such physical lifetimes the long tail of comets with binding energies $x \gsim 0.04$ au$^{-1}$ becomes too small in comparison with the observed one, so we require for it a different source region. We then searched for a suitable source of observed comets that might evolve to HTCs. To this end, we considered the Centaurs coming close to Jupiter's orbit, both inactive and active. We found that Centaurs can produce HTCs, in particular they are trapped very often in the 1:6 MMR with Jupiter (energy $x \simeq 0.058$ au$^{-1}$) which coincides with the clustering in the observed energy distribution of HTCs (Fig. \ref{energies}).\\

It is worth noting that our sample of observed Centaurs contains a few objects in retrograde orbits. To produce both the observed HTCs and JFCs, a flattened source is required but it should contain a small fraction of high-inclination and retrograde comets. A near-ecliptic, prograde population of Centaurs ($i \lsim 25^{\circ}$) will essentially lead to a near-Earth population of bodies with Tisserand parameters $T_J > 2$, typical of JFCs \citep{Levi97}. It is not the purpose of this paper to discuss the ultimate origin of Centaurs, though we can say a few words. A flat source in the outer planetary region such as the Scattered Disk will essentially produce a population of Centaurs in prograde, low-inlination orbits \citep{Disi07}, though it may be possible that some SDOs scattered outwards attain Oort cloud distances where galactic tides switch their orbits from prograde to retrograde \citep{Levi06}. Other possibility is that some Centaurs come from the capture of Oort cloud comets in the outer planetary region as discussed by \citet{Emel98} and \citet{Emel05}. In the end, some HTCs may come from the Oort cloud but through a different dynamical route that direct insertion of Oort cloud comets in the near-Earth or the Jupiter region.\\

\begin{figure}[h]
\resizebox{10cm}{!}{\includegraphics{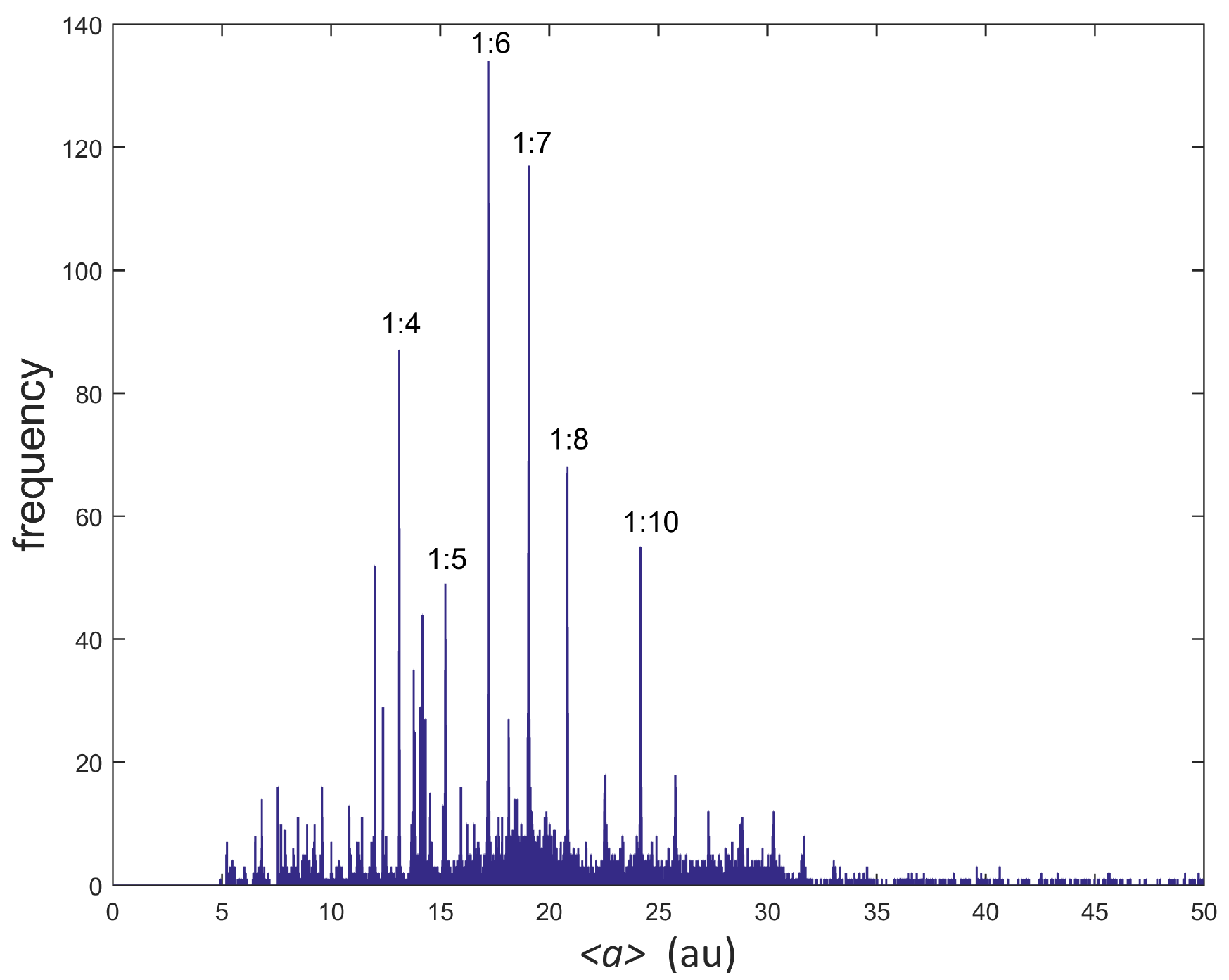}}
\caption{Frequency distribution of the time-average $<a>$ (over 500 yr) for the observed sample of 15 HTCs with $q < 1.3$ au and energies $0.04<x<0.08$ au$^{-1}$ (periods $44.2 < P < 125$ yr) integrated over $10^6$ yr in the past and in the future.}
\label{HTCs}
\end{figure}

In order to check the results found for the simulated HTCs produced from the evolution of the Centaur sample, we also studied the dynamical evolution for 1 Myr to the past and to the future of 15 observed HTCs with $q < 1.3$ au and energies $0.04<x<0.08$ au$^{-1}$ (orbital periods $44.2 < P < 125$ yr). We confirm that HTCs do not diffuse randomly in $a$ but tend to be captured in the 1:6 exterior resonance with Jupiter generating a concentration at $x \sim 0.058$ au$^{-1}$. This peak is clearly shown in Fig. \ref{HTCs} which is a histogram of frequencies of the average $<a>$ over periods of 500 years.\\

Other remarkable feature is that even though we obtain about 4.7\% of LPCs that end up colliding with the Sun for a purely dynamical evolution, this percentage is practically null when we consider appropriate physical lifetimes ($N_{phys}(q=1) \lsim 500$ revolutions). The reason is that LPCs approach the Sun in a smooth way through a multi-step process, that takes hundreds to thousands revolutions, so it is very likely that such comets will disintegrate well before reaching sungrazing states.\\

The inclination distribution of the old LPCs shows clear departures from randomness that can be described as a slight predominance of prograde orbits that is somewhat counterbalaced by an excess of near-ecliptic, retrograde comets ($-1 < \cos{i} < -0.9$). Some predominance of prograde orbits in old LPCs can be explained as a result of dynamical and physical losses of comets random-walking in the energy space \citep{Fern94}, but why is there an excess of near-ecliptic retrograde orbits?. The results shown in Fig. \ref{icent} (bottom, right-hand panel) can provide an explanation to this curious feature. The evolution of Centaurs in the near-Earth region generates a population of fake 'old' LPCs near $180^{\circ}$ that adds to the bona fide old LPCs from the Oort cloud, giving a possible explanation for the observed excess of near-ecliptic retrograde orbits.\\

Another important point to analyze is whether the end states of the evolution of LPCs and Centaurs produce the correct proportions between HTCs and JFCs observed in the near-Earth region ($q < 1.3$ au). Remember (cf. Fig. \ref{tisserand}) that for orbital periods $P < 20$ yr most comets are JFC type (i.e. their Tisserand parameters $T_J > 2$), while the opposite occurs for periods $P > 20$ yr. The populations of HTCs and JFCs in the near-Earth region are of comparable size \citep{Fern05,Emel13}. We find that the evolution of LPCs leads to 99.4\% of HTCs and only 0.6\% of JFCs in the case of a purely dynamical evolution. For a physical lifetime of $N_{phys}(q=1) = 200$ revolutions, the proportions changes somewhat but not dramatically: there are 95\% of HTCs and 5\% of JFCs. If we focus on the comets reaching orbital periods $P < 20$ yr, the proportions are 86.6\% of HTCs and 13.4\% of JFCs. We thus conclude that the evolution of LPCs leads to a overwhelming predominance of HTCs in comparison with JFCs, which is in contradiction with observations.\\

When we consider the end products of the evolution of Centaurs in the near-Earth region the results turn out to be quite different. Thus, a purely dynamical evolution leads to 70.1\% of HTCs and 29.9\% of JFCs. When we only consider those that reach $P < 20$ yr, the proportions are 54.6\% of HTCs and 45.4\% of JFCs. These values are closer to the observed ones, but still not quite consistent. If we consider now a more realistic case of the evolution of Centaurs with a finite physical lifetime ($N_{phys}(q=1) = 500$ revolutions), the proportions become: HTCs (36.6\%) and JFCs (63.4\%), if we restrict the analysis to $P < 20$ yr, the corresponding values are: HTCs (16.1\%) and JFCs (83.9\%). We can see that in this case we can get populations of HTCs and JFCs whose sizes are consistent with the observed ones, in particular we can verify that in this case the great majority of comets with periods $P < 20$ yr are of JFC type.

\section{Conclusions}
\label{conc}

We can sum up the main results of our study in the following points:

\begin{itemize}

\item The simulated dynamical and physical evolution of old LPCs in the near-Earth region can match the observed energy distribution of old LPCs and HTCs only up to energies $x \simeq 0.04$ au$^{-1}$. We get too few comets with energies greater than $\sim 0.04$ au$^{-1}$ as compared to the observed distribution, so a source other than LPCs is required to explain their origin.

\item The best results for the simulated energy, inclination and perihelion distance distributions are obtained for a physical lifetime of about 200-300 revolutions (for a comet with a standard perihelion distance $q=1$ au).

\item The observed long tail of near-Earth comets with binding energies $x \gsim 0.04$ au$^{-1}$ (corresponding to HTCs and JFCs) can be best explained if their immediate source are the Centaurs with perihelia close to Jupiter's orbit. In particular, we are able to explain the clustering of energies around $\sim 0.05 - 0.065$ (orbital periods $P \sim 60-90$ yr) where comets can be trapped in strong MMRs with Jupiter, in particular 1:6 and 1:7.

\item During the transit to HTC orbits, old LPCs are often captured in MMRs with Jupiter, in particular in exterior 1:N resonances. For comets in the near-Earth region reaching semimajor axes $a < 50$ au, we found that al least 12\% of the evolution time is spent in any of these resonances. 

\item A purely dynamical evolution leads to a percentage $\sim 4.7\%$ of LPCs that end up as sungrazers. Yet, no sungrazers were obtained when the number of revolutions is limited by a physical lifetime $\lsim 10^3$ revolutions (for a standard $q=1$ au). This is because it takes from hundreds to thousands of revolutions for a comet to become a sungrazer, starting from a somewhat more distant orbit (say, with a perihelion distance $q=0.2$ au that roughly corresponds to the minimum perihelion distance of observed old LPCs).

\item The boundary between old LPCs (coming from the Oort cloud) and HTCs (coming from the Scattered Disk) is found to lie at $x \sim 0.04$ au$^{-1}$ ($P \sim 125$ yr). Therefore, the traditional limit at $P=200$ yr has no dynamical basis. As shown, comets with energies 0.03-0.04 au$^{-1}$ ($P \sim 125-200$ yr) may correspond to the end tail of the evolution of old LPCs.

\item Most of the comets in the near-Earth region reaching energies $x > 0.04$ au$^{-1}$ from the evolution of LPCs are of HTC type, even among those with orbital periods $P < 20$ yr. Therefore, we find a discrepancy not only with the very small number of comets reaching such short periods, but also that most of them are of HTC type while the observations show that the great majority are JFCs. However a Centaur source produce populations of HTCs and JFCs whose sizes are consistent with observations.

\item Centaurs with perihelia close to Jupiter's orbit are thus very likely the immediate source of HTCs and JFCs. It was beyond the scope of this paper to analyze the provenance of the Centaurs. We can argue that most of them come from the Scattered Disk, being essentially subject to the perturbations of the Jovian planets, with a small fraction that attained Oort cloud distances during their evolution that allow galactic tides to randomize their orbits (thus generating a small percentage of Centaurs in high-inclination and retrograde orbits).  

\end{itemize}

\bigskip

\textbf{Acknowledgments}

We thank the referee, Marc Fouchard, for useful comments on the manuscript. We gratefully acknowledge financial support from Project CSIC Grupo I+D 831725 - Planetary Sciences.

\end{document}